\documentclass[
aps,
twocolumn,
showpacs,
amsmath,
amssymb,
floatfix,
]{revtex4}
\usepackage[dvips]{graphicx}
\usepackage[colorlinks=true]{hyperref}
\usepackage{dcolumn}
\usepackage{bm}
\usepackage{color} 
\usepackage{ulem}

\begin{document}

\title{
Block Lanczos density-matrix renormalization group method for general Anderson impurity models: 
Application to magnetic impurity problems in graphene
}

\author{Tomonori Shirakawa$^{1,2}$} 
\author{Seiji Yunoki$^{1,2,3}$}
\affiliation{$^1$Computational Condensed Matter Physics Laboratory, RIKEN, Wako, Saitama 351-0198, Japan \\
${}^2$Computational Materials Science Research Team, RIKEN Advanced Institute for 
Computational Science (AICS), Kobe, Hyogo 650-0047, Japan \\
${}^3$Computational Quantum Matter Research Team, RIKEN Center for Emergent Matter Science (CEMS), Wako, Saitama 351-0198, Japan }

\date{\today}

\begin{abstract}
We introduce a block Lanczos (BL) recursive technique to construct quasi-one-dimensional models, suitable for
density-matrix renormalization group (DMRG) calculations, from single- as well as multiple-impurity Anderson
models in any spatial dimensions. This new scheme, named BL-DMRG method, allows us to calculate not
only local but also spatially dependent static and dynamical quantities of the ground state for general Anderson
impurity models without losing elaborate geometrical information of the lattice. We show that the BL-DMRG
method can be easily extended to treat a multiorbital Anderson impurity model where not only inter- and
intraorbital Coulomb interactions but also Hund’s coupling and pair hopping interactions are included. We also
show that the symmetry adapted BL bases can be utilized, when it is appropriate, to reduce the computational
cost. As a demonstration, we apply the BL-DMRG method to three different models for graphene with a
structural defect and with a single hydrogen or fluorine absorbed, where a single Anderson impurity is coupled to
conduction electrons in the honeycomb lattice. These models include (i) a single adatom on the honeycomb lattice,
(ii) a substitutional impurity in the honeycomb lattice, and (iii) an effective model for a single carbon vacancy
in graphene. Our analysis of the local dynamical magnetic susceptibility and the local density of states at the
impurity site reveals that, for the particle-hole symmetric case at half-filling of electron density, the ground state
of model (i) behaves as an isolated magnetic impurity with no Kondo screening, while the ground state of the
other two models forms a spin-singlet state where the impurity moment is screened by the conduction electrons.
We also calculate the real-space dependence of the spin-spin correlation functions between the impurity site and
the conduction sites for these three models. Our results clearly show that, reflecting the presence or absence of
unscreened magnetic moment at the impurity site, the spin-spin correlation functions decay as $\propto r^{-3}$, differently
from the noninteracting limit ($\propto r^{-2}$), for model (i) and as $\propto r^{-4}$, exactly the same as the noninteracting limit,
for models (ii) and (iii) in the asymptotic $r$, where $r$ is the distance between the impurity site and the conduction
site. Finally, based on our results, we shed light on recent experiments on graphene where the formation of local
magnetic moments as well as the Kondo-like behavior have been observed.
\end{abstract}


\maketitle

\section{\label{sec:introduction}Introduction}

Recently, magnetic properties of graphene monolayers have attracted much 
attention~\cite{novoselov1,novoselov2,geim,neto}. 
Because of the characteristic electronic band structure with Dirac-like linear dispersions near the Fermi level ($E_{\rm F}$) 
and the resulting ``V-shape'' electronic density of states, $\rho(\omega)\sim|\omega|$, at $E_{\rm F}$~\cite{novoselov2,neto,wallace}, 
a unique electronic and magnetic behavior is expected in graphene. Among many, very recent experiments have 
revealed that hydrogen or fluorine adatoms as well as vacancies in graphene can induce magnetic moments with spin 1/2 per adatom or 
vacancy~\cite{nair,maccreary}, although pristine graphene is diamagnetic.  
Moreover, some experiments have observed the Kondo-like signature in the temperature dependence of 
the resistivity when the vacancies are introduced in graphene~\cite{chen}, even though the other experiments have found otherwise~\cite{nair}. 

The early theoretical studies have considered a single magnetic impurity coupled to the graphene conduction electrons 
and found that the magnetic impurity is completely isolated, i.e., unscreened by the conduction electrons, when the model 
preserves the particle-hole symmetry, while the magnetic moment can be screened 
when the model is strongly particle-hole asymmetric~\cite{withoff,bulla2,gonzalez-baxton,ingersent,fritz2,vojta1,fritz1}.
These theoretical results appear to contradict the experimental observation reported in Ref.~\cite{chen}, where 
the Kondo temperature is found to be symmetric with respect to the applied gate voltage, 
which changes the chemical potential of the conduction electrons. 
Therefore, the experiments indicate that the graphene with vacancies is close to the particle-hole symmetric point. 
However, the early theoretical studies predict no Kondo screening in this limit~\cite{withoff,bulla2,gonzalez-baxton,ingersent,fritz2,vojta1,fritz1}.

Motivated by these experiments~\cite{nair,maccreary,chen}, several models have been recently proposed to explain the origin of 
magnetic moment and the possible Kondo-like effect in graphene with structural defects or 
adatoms~\cite{fritz1,sengupta,vojta2,uchoa,kanao,nanda,cazalilla}. 
One of the possible explanations of the emergent magnetic moment in graphene with a structural defect is 
due to the partially filled dangling bonds of $sp^2$ orbital on carbon atoms surrounding the 
vacancy~\cite{kanao,nanda,cazalilla,barbary,yazyev,amara,palacios}. 
It has been also pointed out that the scattering of defects drastically changes the 
electronic structures of $\pi$-band and produces the logarithmic divergence at $E_{\rm F}$ 
in the local density of states at the vicinity of defects~\cite{cazalilla,vmpereira,peres}. 
Therefore a nonperturbative real-space theoretical approach that can incorporate the elaborate lattice geometry is highly desirable to 
understand the magnetic properties of graphene with structural defects or adatoms. 

The interests in the real-space aspects of magnetic impurities is not only to study geometrically different lattice structures 
of various systems but also to directly capture the real-space nature of the ground state, e.g., 
the spatial distribution of ``Kondo cloud'' in the Kondo singlet state~\cite{affleck,mitchell}. 
Indeed, as compared to the thermodynamics and the transport properties, 
the real-space nature of Kondo problem has been much less studied both experimentally and theoretically. 
However, very recently, 
scanning tunneling spectroscopy experiments have successfully observed the long-range Kondo signature for single magnetic 
atoms of Fe and Co in a Cu(110) surface and found that 
the Kondo cloud seems rather spatially extended away from the magnetic atoms~\cite{prueser}. 
The recent experimental progress further encourages us to study the magnetic impurity problems and the Kondo physics in real space. 

Theoretically, on the other hand, it is still difficult to study the real-space properties simply because the analytical approaches 
available are rather limited and also because even the most powerful and well accepted numerical method for magnetic impurity problems, i.e., 
numerical renormalization group (NRG) method~\cite{wilson,krishna-murthy,bulla1}, 
can not treat the real-space dependence directly, where the 
high-energy scales are integrated out by using logarithmic discretization of energy. 
Quantum Monte Carlo (QMC) methods can calculate spatially dependent quantities~\cite{hirsch,gubernatis,gull}. 
However, the QMC calculations often suffer the negative sign problem at low temperatures 
and can not be applied to general Anderson impurity models.  

In the last decade, the density-matrix renormalization group (DMRG) method~\cite{white1,white2,schollwock,hallberg1} has been 
successfully used to investigate limited properties of single- and multiple-impurity Anderson models. 
For example, the DMRG method has been applied to the single- and two-impurity Anderson/Kondo models 
in one dimension to study correlation effects in the conduction sites~\cite{hallberg2,costamagna,tiegel} and 
to evaluate the Kondo screening length~\cite{sorensen,hand,holzner1}. 
Moreover, the DMRG method has been employed to address single- and multiple-impurity Anderson models in more 
than one dimension~\cite{nishimoto1,nishimoto2,peter} and also applied as an impurity solver for the dynamical mean-field 
theory (DMFT)~\cite{garcia,nishimoto3}. 
However, these approaches encounter difficulties in calculating the spatially dependent quantities such as spin-spin 
correlation functions. 
Therefore, it is highly desired to develop new numerical methods, which can compute directly various physical quantities 
in real space in any spatial dimensions. 

To overcome the difficulties, here we introduce a block Lanczos (BL) recursive technique, which constructs, 
without losing any geometrical information of the lattice, quasi-one-dimensional (Q1D) models, suitable for DMRG 
calculations, from single- as well as multiple-impurity Anderson models in any spatial dimensions. 
This new approach, named BL-DMRG method, enables us to calculate various physical quantities directly in real space, including 
both static and dynamical quantities, with high accuracy. 
Thus the BL-DMRG method is in sharp contrast to the NRG method since the NRG method 
has a severe limitation in calculating the spatially dependent quantities 
because the logarithmic discretization in energy space has to be introduced to 
construct the Wilson chain~\cite{bulla1}.
The BL-DMRG method is also superior to the QMC methods 
because the BL-DMRG method can be easily extended to a more involved impurity model such as a multiorbital single-impurity 
Anderson model where intra- and interorbital Coulomb interactions as well as Hund's coupling and pair hopping interactions are 
included. Therefore, the BL-DMRG method has potential as a promising impurity solver of DMFT 
for multiorbital Hubbard models~\cite{modmft}.

To demonstrate the BL-DMRG method, we apply this method to three different models for graphene with a structural defect and 
with a single absorbed atom, 
where a single Anderson impurity is coupled to the conduction electrons in the honeycomb lattice. 
These models include (i) an Anderson impurity absorbed on the honeycomb lattice (model I), 
(ii) a substitutional Anderson impurity in the honeycomb lattice (model II), and (iii) 
an effective model for a single carbon vacancy in graphene (model III). 
Our results of the local magnetic susceptibility and the local density of states at the impurity site 
reveal that, for the particle-hole symmetric case at half-filling of electron density, the ground state of model I behaves as an 
isolated magnetic impurity with no Kondo screening, while the ground state of models II and III forms a spin 
singlet state where the impurity moment is screened by the conduction electrons. 
To understand the real-space spin distribution of the conduction electrons around the impurity, 
we subsequently calculate the spin-spin correlation functions between the impurity site 
and the conduction sites and find a qualitatively different asymptotic behavior when compared with the noninteracting limit, 
which results from the different screening characteristics: 
the spin-spin correlations decay as $\propto r^{-3}$, different from the noninteracting limit ($\propto r^{-2}$), for model I 
and as $\propto r^{-4}$, exactly the same as the noninteracting limit, for models II and III.  
We also discuss the relevance of our results to the recent experiments on graphene with 
structural defects and with hydrogen or fluorine adatoms where the formation of local magnetic moments and the Kondo-like 
behavior have been observed~\cite{nair,maccreary,chen}. 

The rest of this paper is organized as follows. First, we introduce the BL-DMRG method for 
general Anderson impurity models and describe the details in Sec.~\ref{sec:method}. 
The BL recursive technique is employed to construct Q1D models from general Anderson impurity models in any spatial 
dimensions and for any lattice geometry without losing the structural information in Sec.~\ref{sec:q1dmap}. 
To optimize the DMRG calculations for Q1D models constructed by the BL recursive technique, symmetrization schemes of the BL bases 
are described in Sec.~\ref{sec:sym}. 
The numerical technique to calculate spatially dependent quantities in real space away from the impurity site is 
explained in Sec.~\ref{sec:orbs}. The extension of the BL-DMRG method and the symmetry adapted BL bases for 
a multiorbital single-impurity Anderson model are provided in Sec.~\ref{sec:moam}. 

The BL-DMRG method is then demonstrated in Sec.~\ref{sec:application} for single-impurity Anderson models. 
The three different single-impurity Anderson models for graphene with a structural defect and with a single adatom 
are introduced in Sec.~\ref{sec:model}. After briefly explaining the numerical details of the calculations for these models in 
Sec.~\ref{sec:numerical}, 
the nature of the ground state is examined by calculating the local magnetic susceptibility 
at the impurity site in Sec.~\ref{sec:dssf} and the local electronic density of states at the impurity site in Sec.~\ref{sec:ldos}. 
The spin-spin correlation functions between the impurity site and the conduction sites 
are evaluated in Sec.~\ref{sec:sscf}. The relevance of our results to the recent 
experiments on adatoms or vacancies in graphene is discussed in Sec.~\ref{sec:summary}. 
The possible further extension of the BL-DMRG method is also briefly discussed. 
The detailed derivation of the hybridization function for general Anderson impurity models 
is described in Appendix~\ref{app:hybfunc}. 

\section{\label{sec:method} BL-DMRG Method}

In this section, we introduce the BL-DMRG method for general Anderson impurity models in any spatial dimensions and for any 
lattice geometry. To this end, first we describe in Sec.~\ref{sec:q1dmap} the BL recursive technique 
which enables us to transform exactly a general Anderson impurity model to a Q1D model without losing any geometrical information 
of the lattice. 
Once a Q1D model is constructed, we can use the DMRG method to calculate both static and dynamical quantities with 
extremely high accuracy. 

We then describe in Sec.~\ref{sec:sym} two schemes to reduction the computational cost for DMRG 
calculations. One is to utilize the lattice symmetry of the models to construct the symmetry adapted BL bases, 
which is similar to the one introduced in NRG calculations for multi-impurity problems~\cite{jones,silva,borda}. 
The other is to use spin degrees of freedom to reduce the dimensions of the local Hilbert space, 
which can be applied to more general cases even if the models do not possess appropriately high lattice symmetry. 
The BL-DMRG procedure to calculate spatially dependent quantities such as spin-spin correlation functions 
is also explained in Sec.~\ref{sec:orbs}. Finally, the extension to a multiorbital single-impurity Anderson model is briefly 
discussed in Sec.~\ref{sec:moam}. 

It should be emphasized that, although the BL-DMRG method shares some similarity with the NRG method~\cite{borda}, 
the BL-DMRG method can be readily extend to more general models, 
one example discussed in Sec.~\ref{sec:moam}, 
and has significant advantages in calculating spatially dependent quantities and also in the computational cost by using the 
symmetry adapted BL bases. 
We should also note that, very recently, the direct application of a standard Lanczos technique to single-impurity 
Anderson and Kondo models~\cite{brusser} as well as its extension to a two-impurity Kondo model~\cite{allerdt} 
have been proposed for DMRG calculations, 
which is somewhat similar to the BL-DMRG method introduced in this paper. 
However, we emphasize that the use of BL recursive technique in the BL-DMRG method significantly enlarges the applicability of 
DMRG calculations not only to more general multiorbital single- or multiple-impurity Anderson models but also to the calculations 
of spatially dependent quantities. 
Moreover, as described in Appendix~\ref{app:hybfunc}, the BL bases representation of the hybridization function for general Anderson impurity 
models further enlarges the usefulness of the BL recursive technique for other 
numerical methods such as the QMC methods and the NRG method.

\subsection{\label{sec:q1dmap} Q1D map of a general Anderson impurity model: a BL recursive technique}

We consider a general Anderson impurity model described by the following Hamiltonian: 
\begin{equation}
\mathcal{H}_{\rm AIM}= \mathcal{H}_{c} + \mathcal{H}_{d} + \mathcal{H}_{V}  + \mathcal{H}_{U}, \label{eq:generalham}
\end{equation}
where 
\begin{eqnarray}
\mathcal{H}_{c} &=& \sum_{n,n^{\prime}} \sum_{\sigma} \epsilon_{n,n^{\prime}}^c 
c_{n,\sigma}^{\dagger} c_{n^{\prime},\sigma}, \\
\mathcal{H}_{d} &=& \sum_{m,m^{\prime}} \sum_{\sigma} \epsilon_{m,m^{\prime}}^d d_{m,\sigma}^{\dagger} d_{m^{\prime},\sigma}, \\
\mathcal{H}_{V} &=& \sum_{m,n} \sum_{\sigma} ( V_{m,n} 
d^{\dagger}_{m,\sigma} c_{n,\sigma} + {\rm h.c.}), 
\end{eqnarray}
and 
\begin{eqnarray}
\mathcal{H}_{U} &=& \sum_{ m_1,\cdots,m_4} \sum_{\sigma_1,\cdots,\sigma_4}
 U_{m_1m_2;m_3m_4}^{\sigma_1\sigma_2;\sigma_3\sigma_4} 
d_{m_1, \sigma_1}^{\dagger} d_{m_2, \sigma_2}^{\dagger} \nonumber \\ 
&&\quad\quad\quad\quad\quad\quad\quad\quad\quad\quad\quad    \times d_{m_3, \sigma_3} d_{m_4, \sigma_4}. \label{eq:generalu}
\end{eqnarray}
Here, $c_{n,\sigma}^{\dagger}$ ($c_{n,\sigma}$) is the creation (annihilation) operator 
of an electron at site (or orbital) $n$ ($=1,2,\cdots,N$) with spin $\sigma$ ($= \uparrow, \downarrow$) 
in the conduction sites (or bands) 
and $d_{m,\sigma}^{\dagger}$ ($d_{m,\sigma}$) is the creation (annihilation) operator 
of an electron at impurity site $i$ ($=1,2,\cdots,N_i$) and orbital $\alpha$ ($=1,2,\cdots,N_d$), denoted by 
$m=(i,\alpha)$ ($=1,2,\cdots,M$, where $M=N_iN_d$) for simplicity, with spin $\sigma$. The individual terms, 
$\mathcal{H}_c$, $\mathcal{H}_d$, $\mathcal{H}_V$, and $\mathcal{H}_U$, describe the one-body part of the conduction 
sites (or bands), 
the one-body part of the impurity sites, the hybridization between the impurity sites and the conduction sites (or bands), 
and the two-body Coulomb interaction part of the impurity sites, respectively. 
Notice that this Hamiltonian includes a wide range of Anderson impurity models, ranging from 
the simplest single-orbital single-impurity Anderson model ($N_i=N_d=1$) to a more complex multiorbital multiple-impurity Anderson 
model ($N_i,N_d>1$).
Notice also that neither the spatial dimensions nor the lattice geometry is assumed for $\mathcal{H}_{\rm AIM}$. 

We shall now show that the general Anderson impurity model $\mathcal{H}_{\rm AIM}$ given in Eq.~(\ref{eq:generalham}) 
can be mapped onto a Q1D ladder-like model, for which the DMRG method is applied with high accuracy. 
This Q1D mapping can be achieved exactly without losing any geometrical information of the lattice by using the BL recursive technique, 
which is a straightforward extension of the basic Lanczos recursive procedure~\cite{blt}. 
To simplify the formulation, let us first introduce the vector representation of fermion operators: 
\begin{equation}\label{eq:cop}
{\bm c}_{\sigma}^{\dagger}=(d^{\dagger}_{1,\sigma}, d^{\dagger}_{2,\sigma}, 
\cdots, d^{\dagger}_{M,\sigma}, 
c^{\dagger}_{1,\sigma},c^{\dagger}_{2,\sigma},\cdots,c^{\dagger}_{N,\sigma}).
\end{equation} 
Then, the one-body part of the Hamiltonian, $\mathcal{H}_0 = \mathcal{H}_c + \mathcal{H}_d + \mathcal{H}_V$, 
in Eq.~(\ref{eq:generalham}) can be represented as
\begin{eqnarray}
\mathcal{H}_0 = \sum_{\sigma} {\bm c}^{\dagger}_{\sigma} \hat{H}_0 {\bm c}_{\sigma} \label{eq:one-body}
\end{eqnarray}
with 
\begin{eqnarray}
\hat{H}_0 = \left( 
\begin{array}{cc}
\hat{H}_d & \hat{V} \\
\hat{V}^{\dagger} & \hat{H}_c \\
\end{array}
\right), \label{eq:onebodyham}
\end{eqnarray}
where $\hat{H}_d$, $\hat{H}_c$, and $\hat{V}$ are 
$M \times M$, $N \times N$, and $M \times N$ matrices with matrix elements 
$(\hat{H}_d)_{m,m^{\prime}}=\epsilon_{m,m^{\prime}}^d$, 
$(\hat{H}_c)_{n,n^{\prime}}=\epsilon_{n,n^{\prime}}^c$, and 
$(\hat{V})_{m,n}=V_{m,n}$, respectively. 

Next, let us construct the following matrix $\hat{P}_1$ composed of $M$ different vectors ${\bm e}_m$: 
\begin{eqnarray}
\hat{P}_1 = \left( 
{\bm e}_1, {\bm e}_2, \cdots, {\bm e}_M
\right), \label{eq:blinit}
\end{eqnarray}
where ${\bm e}_{m}$ is a $(N+M)$-dimensional column unit vector with its element 
$({\bm e}_m)_{n}=\delta_{m,n}$ and thus $\hat{P}_1$ is a $(N+M) \times M$ matrix. 
Using $\hat{P}_1$ as  the initial BL bases, 
the Krylov subspace of $\hat{H}_0$ is spanned with the BL bases $\hat{P}_{l+1}$ ($l=1,2,\cdots$) 
generated through the three-term recurrences,  
\begin{eqnarray}
\hat{P}_{l+1} \hat{T}_{l}^{\dagger} = \hat{H}_0 \hat{P}_l - \hat{P}_l \hat{E}_l - \hat{P}_{l-1} \hat{T}_{l-1}, \label{eq:bliter}
\end{eqnarray}
where $\hat E_l=\hat P_l^\dag \hat H_0 \hat P_l$, $\hat{P}_0=0$, and $ \hat{T}_0= 0$. The left hand side of Eq.~(\ref{eq:bliter}) is obtained 
with a QR factorization of the $(N+M)\times M$ matrix in the right hand side of Eq.~(\ref{eq:bliter}). 
Thus, $\hat P_{l+1}$ is a column orthogonal 
$(N+M) \times M$ matrix and $\hat T_{l}$ is a lower triangular $M \times M$ matrix with $(\hat{T}_l)_{m,m^{\prime}}=0$ for
 $m < m^{\prime}$, i.e.,  
\begin{eqnarray}
\hat{T}_l = \left( 
\begin{array}{ccccc}
T_{11}^{(l)} & 0 & 0 & \cdots & 0 \\
T_{21}^{(l)} & T_{22}^{(l)} & 0 & \cdots & 0 \\
T_{31}^{(l)} & T_{32}^{(l)} & T_{33}^{(l)} & \cdots & 0\\
\vdots & \vdots & \vdots & \ddots & \vdots \\
T_{M1}^{(l)} & T_{M2}^{(l)} & T_{M3}^{(l)} & \cdots & T_{MM}^{(l)} \\
\end{array}
\right). 
\end{eqnarray}
 After repeating this procedure, 
the BL bases are constructed 
and they are gathered in the $(N+M) \times (N+M)$ matrix $\hat{P}$:  
\begin{eqnarray}
\hat{P} = \left( \hat{P}_1, \hat{P}_2, \hat{P}_3, \cdots \right), 
\label{eq:pdef}
\end{eqnarray}
which satisfies $\hat{P}^{\dagger}\hat{P}= \hat{P} \hat{P}^{\dagger}=\hat{I}$ where $\hat{I}$ is the unit matrix. 
It should be noted here that, in practical calculations where $N$ is large, we usually terminate the BL iteration 
after the $L$th iteration, 
for which $\hat{P}$ is a rectangular $(N+M)\times (LM)$ matrix satisfying only $\hat{P}^{\dagger} \hat{P} = \hat{I}$ 
but not $\hat{P} \hat{P}^{\dagger} = \hat{I}$, in general. 

With this unitary matrix $\hat P$, the one-body part $\mathcal H_0$ of the Hamiltonian can now be block-tridiagonalized, 
\begin{eqnarray}
\mathcal{H}_0 = \sum_{\sigma} {\bm c}^{\dagger}_{\sigma} \hat{P} 
\hat{P}^{\dagger} \hat{H}_0 \hat{P} \hat{P}^{\dagger} {\bm c}_{\sigma} 
= \sum_{\sigma} {\bm a}_{\sigma}^{\dagger} \hat{H}_0^{\rm BL}
{\bm a}_{\sigma} \label{eq:blbloc}
\end{eqnarray}
with 
\begin{eqnarray}
\hat{H}_0^{\rm BL} = \left(
\begin{array}{ccccc}
\hat{E}_1 & \hat{T}_1 & 0 & 0 & \cdots \\
\hat{T}_1^{\dagger} & \hat{E}_2 & \hat{T}_2 & 0 & \cdots \\
0 & \hat{T}_2^{\dagger} & \hat{E}_3 & \hat{T}_3 & \cdots \\
0 & 0 & \hat{T}_3^{\dagger} & \hat{E}_4 & \ddots \\
\vdots & \vdots & \vdots & \ddots & \ddots \\
\end{array}
\right) \label{eq:modonebodyham}
\end{eqnarray}
and ${\bm a}_{\sigma}=\hat{P}^{\dagger} {\bm c}_{\sigma}$~\cite{note5}. 
Note that $\hat{T}_l$ is a lower triangular $M \times M$ matrix 
and thus $\hat{H}_0^{\rm BL}$ has the bandwidth of $2M+1$. 
Hereafter, we will use the following convention for the indices of ${\bm a}_{\sigma}^{\dagger}$: 
\begin{equation}
{\bm a}_{\sigma}^{\dagger}=(a_{1,1,\sigma}^{\dagger}, a_{1,2,\sigma}^{\dagger},\cdots , 
a_{1,M,\sigma}^{\dagger},a_{2,1,\sigma}^{\dagger},\cdots,a_{l,m,\sigma}^{\dagger},\cdots),
\label{a_op}
\end{equation} 
and thus 
\begin{eqnarray}
({\bm a}_{\sigma})_{lm} & = & 
  \sum_{n=1}^{N+M}\left(  \hat P^\dag_l \right)_{m,n} ({\bm c}_{\sigma})_n,  \label{eq:ut_a}\\
 ({\bm c}_{\sigma})_{n} & = & 
  \sum_{l=1}^{N/M+1}\sum_{m=1}^M\left(  \hat P_l \right)_{n,m} ({\bm a}_{\sigma})_{lm}, \label{eq:ut_c}
\end{eqnarray}
where $({\bm a}_{\sigma})_{lm}=a_{l,m,\sigma}$ and ${\bm c}_{\sigma}$ is given in Eq.~(\ref{eq:cop})~\cite{note5}. 
It is important to notice that, because of the special choice of the initial BL bases $\hat{P}_1$ in Eq.~(\ref{eq:blinit}), 
the new fermionic operators representing the impurity sites 
remain unchanged, i.e., 
\begin{equation}
(d_{1,\sigma}^\dag, d_{2,\sigma}^\dag, \cdots, d_{M,\sigma}^\dag) 
= (a_{1,1,\sigma}^{\dagger}, a_{1,2,\sigma}^{\dagger},\cdots , a_{1,M,\sigma}^{\dagger}).\label{eq:d_a}
\end{equation}  
Therefore, the two-body part $\mathcal H_U$ of the Hamiltonian is exactly in the same form for 
the new fermionic operator ${\bm a}_{\sigma}$. This is the crucial point for the exact mapping of any Anderson impurity model 
onto a Q1D model. 

\begin{figure}[htbp]
\begin{center}
\includegraphics[width=80mm]{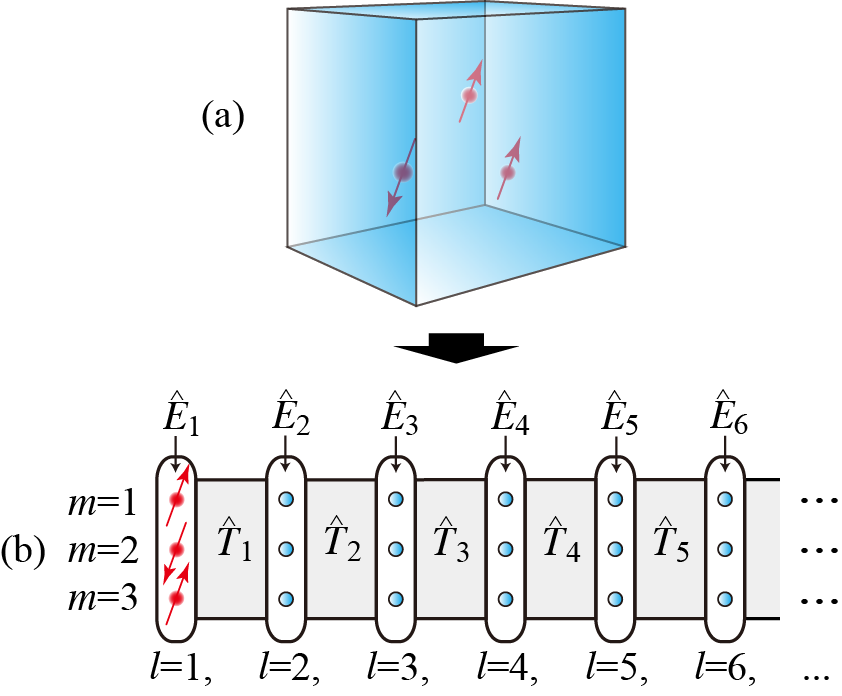}
\caption{(Color online) Schematic representation of the Q1D mapping for a single-orbital three-impurity Anderson model 
($N_i=3$, $N_d=1$, and $M=3$) in three dimensions. Using the BL recursive technique, the Anderson impurity 
model $\mathcal H_{\rm AIM}$ is transformed exactly to a semi-infinite three-leg ladder model $\mathcal H_{\rm AIM}^{\rm Q1D}$ 
without loosing any geometrical information of the lattice. 
The conduction sites (or orbitals) are indicated by a blue cube in (a). 
The Coulomb interaction term ${\mathcal H}_U$ is active only at the impurity sites (denoted by 
red circles with arrows) both in (a) and (b). 
Blue circles without arrows in (b) represent the ``ripple'' sites (i.e., BL bases) generated by the BL recursive procedure. 
The indices $l$ and $m$ in (b), representing the site position of the resulting three-leg ladder model along the leg and rung directions, 
respectively, correspond to the ones in the new fermionic operator $({\bm a}_{\sigma})_{lm}=a_{l,m,\sigma}$ in 
Eq.~(\ref{a_op}) and used to describe $\mathcal H_{\rm AIM}^{\rm Q1D}$ in Eq.~(\ref{eq:hamlad}). 
The ``on-site potential'' and ``nearest-neighbor hopping'' matrices, $\hat E_l$ and $\hat T_l$, respectively, in Eq.~(\ref{eq:hamladt}) 
are also indicated in (b). 
}
\label{fig:1dmap}
\end{center}
\end{figure}

It is now apparent that, using the BL recursive technique introduced above, a general Anderson impurity model $\mathcal H_{\rm AIM}$ 
in any spatial dimensions and for any lattice geometry can be mapped exactly onto a Q1D model, i.e., 
a semi-infinite $M$-leg ladder model, described by the following Hamiltonian: 
\begin{equation}
\mathcal{H}_{\rm AIM}^{\rm Q1D} = \sum_\sigma\mathcal{H}_{0,\sigma}^{\rm Q1D} + \mathcal{H}_{U}, \label{eq:hamlad} 
\end{equation}
where 
\begin{eqnarray}
&{}&\mathcal{H}_{0,\sigma}^{\rm Q1D} = \sum_{l=1}^L \sum_{m,m'=1}^{M} (\hat{E}_l)_{m,m'} a_{l,m,\sigma}^{\dagger} a_{l,m',\sigma} 
\nonumber \\ 
&{}&\ + \sum_{l=1}^{L-1}\sum_{m,m'=1}^M  \left ( (\hat{T}_{l})_{m,m'} a_{l,m,\sigma}^{\dagger} a_{l+1,m',\sigma} + {\rm H.c.}\right) 
\label{eq:hamladt}  
\end{eqnarray}
and $\mathcal H_{U}$ is the same two-body Coulomb interaction term given in Eq.~(\ref{eq:generalu}). 
Notice that the index $l$ in Eq.~(\ref{eq:hamladt}) corresponds to the one in the BL iteration in Eq.~(\ref{eq:bliter}), which 
is terminated at the $L$-th iteration~\cite{note2}. 
The schematic representation of the Q1D mapping for an Anderson impurity model with $N_i=3$ and 
$N_d=1$ (thus $M=3$) is shown in Fig.~\ref{fig:1dmap}. 

It should be noticed that the resulting Q1D ladder model $\mathcal{H}_{\rm AIM}^{\rm Q1D}$ in the BL bases 
with $L$ sites along the leg direction therefore represents the original system $\mathcal H_{\rm AIM}$ 
with approximately at least $\pi L^2$ and $4\pi L^3/3$ conduction sites (or orbitals) in two and three spatial dimensions, respectively 
(see, e.g., Figs.~\ref{fig:ripple}~\ref{fig:ripple_sym}, and Fig.~\ref{fig:dorbit}). 
This implies that, as long as the impurity 
properties are concerned, the BL-DMRG method can treat quite large systems with reasonable computational cost for a wide variety of 
Anderson impurity models.

\subsection{\label{sec:sym}Symmetrization of BL bases}

\begin{figure}[htbp]
\begin{center}
\includegraphics[width=65mm]{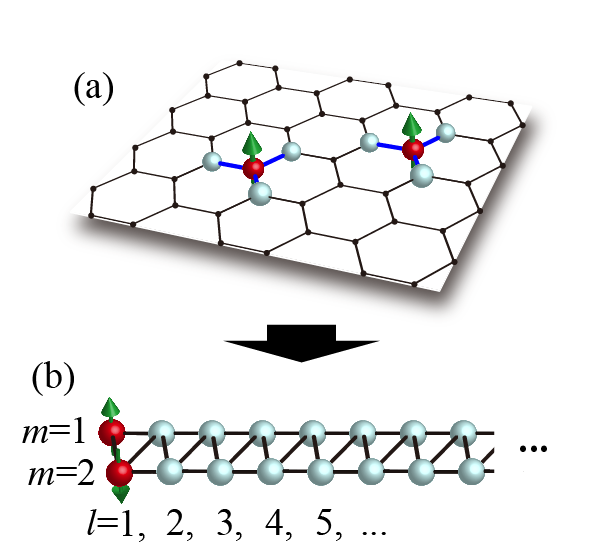}
\caption{(Color online) (a) Schematic representation of a two-impurity Wolff model on the honeycomb lattice described by 
$\mathcal H_{\rm WM}$ in Eq.~(\ref{eq:ham}) and (b) the semi-infinite Q1D ladder model obtained by the BL recursive technique. 
Solid red spheres with green arrows indicate the impurity sites. 
White spheres next to the impurity sites in (a) represent the first ripple states generated by the second ($l=2$) BL iteration 
in Eq.~(\ref{eq:bliter}). The hybridization (with its strength $V$) between the impurity sites and the conduction sites are indicated 
by bold blue lines in (a). 
The indices $l$ and $m$ in (b), representing the site position of the resulting Q1D ladder model along the leg and rung directions, 
respectively, correspond to the ones in the new fermionic operator $({\bm a}_{\sigma})_{lm}=a_{l,m,\sigma}$ used in 
Eq.~(\ref{eq:hamq1d}). 
}
\label{fig:wmodel}
\end{center}
\end{figure}

In this section, we shall describe how the symmetry of Hamiltonian can be used to further simplify the Q1D model 
constructed by the BL recursive technique. This is best explained by taking a specific model. Therefore, as an example, 
we now consider a two-impurity Wolff model on the honeycomb lattice~\cite{wolff}, 
where two conduction sites on the honeycomb lattice are replaced by two impurity sites, 
as schematically shown in Fig.~\ref{fig:wmodel}(a). The Hamiltonian of the Wolff model is 
\begin{equation}
\mathcal{H}_{\rm WM} = \mathcal{H}_c^{\rm WM} + \mathcal{H}_V^{\rm WM} + \mathcal{H}_U^{\rm WM}, \label{eq:ham} 
\end{equation}
where 
\begin{eqnarray}
\mathcal{H}_c^{\rm WM} &=& -t \sum_{\left< {\bf r}, {\bf r}^{\prime}\right>}\sum_\sigma 
\left( c_{{\bf r},\sigma}^{\dagger} c_{{\bf r}^{\prime},\sigma} + \rm{H.c.} \right), \label{eq:hamkin} \\
\mathcal{H}_V^{\rm WM} &=& V \sum_{\left< {\bf r}, {\bf r}^{\prime}\right>^{\prime}}\sum_\sigma 
\left(c_{{\bf r},\sigma}^{\dagger} c_{{\bf r}^{\prime},\sigma} + \rm{H.c.} \right), \label{eq:hamhyb} 
\end{eqnarray}
and
\begin{equation}
\mathcal{H}_U^{\rm WM} = \sum_{{\bf r} \in {\rm Imp.}} U_{\bf r}
\left( n_{{\bf r},\uparrow} - 1/2 \right) 
\left( n_{{\bf r},\downarrow} - 1/2 \right). \label{eq:hamimp}
\end{equation}
Here, $c_{{\bf r},\sigma}^{\dagger}$ ($c_{{\bf r},\sigma}$) is the electron creation (annihilation) operator 
at site ${\bf r}$ on the honeycomb lattice with spin $\sigma (= \uparrow,\downarrow)$ and  
$n_{{\bf r},\sigma}=c_{{\bf r},\sigma}^{\dagger}c_{{\bf r},\sigma}$. 
The sum $\left< {\bf r}, {\bf r}^{\prime}\right>$ in $\mathcal{H}_c^{\rm WM} $ runs over all pairs of nearest-neighbor sites 
except for the ones connecting to the impurity sites, whereas the sum $\left< {\bf r},{\bf r}^{\prime}\right>^{\prime}$ 
in $\mathcal{H}_V^{\rm WM}$ runs over all pairs of nearest-neighbor sites only involving the impurity sites. 
${\mathcal H}_U^{\rm WM}$ represents the on-site Coulomb interaction at the impurity sites with site dependent interaction $U_{\bf r}$, 
and the sum in ${\mathcal H}_U^{\rm WM}$ includes only the impurity sites. 
The two-impurity Wolff model described by $\mathcal{H}_{\rm WM}$ is a special case of the general Anderson impurity model 
$\mathcal{H}_{\rm AIM}$ in Eq.~(\ref{eq:generalham}) with $N_i=2$, $N_d=1$, and $M = 2$.

Applying the BL recursive technique described in Sec.~\ref{sec:q1dmap}, 
we can readily show that the two-impurity Wolff model $\mathcal{H}_{\rm WM}$ is mapped exactly to the Q1D ladder model 
described by the following Hamiltonian: 
\begin{eqnarray}
&{}& \mathcal{H}^{\rm Q1D}_{\rm WM} = \sum_{m=1}^2 U_{{\bf r}_m} ( n_{1,m,\uparrow} - 1/2 ) ( n_{1,m,\downarrow} -1/2 ) \nonumber \\
&{}&  + \sum_{l=1}^L  \sum_{m,m^{\prime}=1}^2 \sum_\sigma \epsilon_{mm^{\prime}}^l a_{l,m,\sigma}^{\dagger} a_{l,m^{\prime},\sigma} \nonumber \\
&{}&  + \sum_{l=1}^{L-1}  \sum_{m,m^{\prime}=1}^2 \sum_\sigma t_{mm^{\prime}}^l ( a_{l,m,\sigma}^{\dagger} a_{l+1,m^{\prime},\sigma} + {\rm H.c.}), 
\label{eq:hamq1d}
\end{eqnarray}
where ${\bf r}_m$ represents the position of $m$th impurity site, $\epsilon_{mm^{\prime}}^l=(\hat E_l)_{m,m'}$, 
$t_{mm^{\prime}}^l=(\hat T_l)_{m,m'}$, and 
$n_{1,m,\sigma}=a_{1,m,\sigma}^{\dagger}a_{1,m,\sigma}=c_{{\bf r}_m,\sigma}^\dag c_{{\bf r}_m,\sigma}$. 
A schematic representation of this Q1D ladder model is shown in Fig.~\ref{fig:wmodel}(b).

\begin{figure}[htbp]
\begin{center}
\includegraphics[width=0.95\hsize]{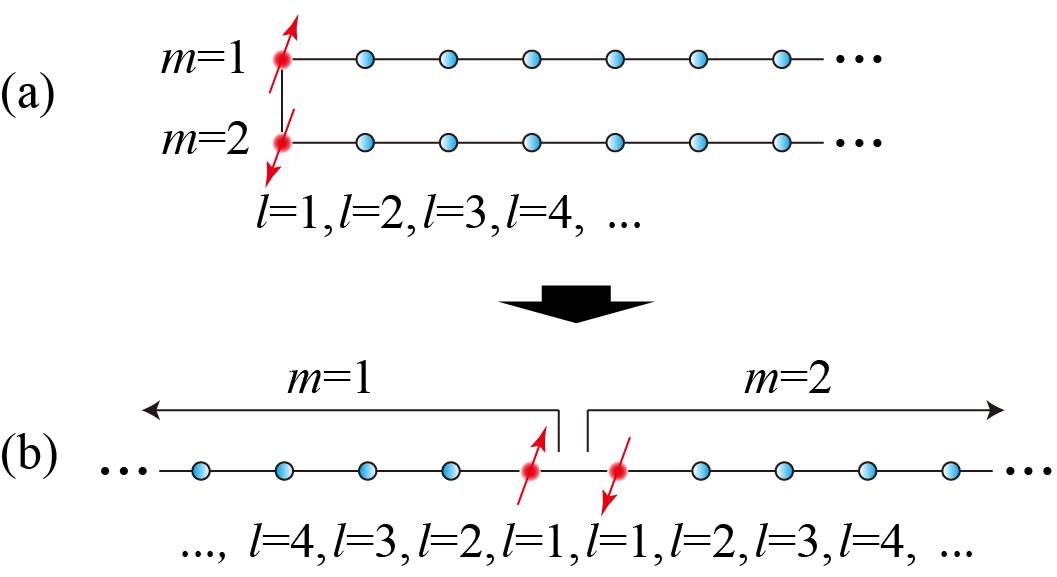}
\caption{(Color online) Schematic representation of a pure one-dimensional model mapped from the two-impurity Wolff model on the honeycomb lattice shown in Fig.~\ref{fig:wmodel} (a). 
Using the symmetry adapted BL bases, (a) the hopping terms between the legs are completely eliminated except for the impurity sites 
and thus (b) the ladder model is further simplified to a pure one-dimensional model. 
Red solid circles with arrow indicate the impurity sites and blue circles without arrows represent the symmetry adapted BL bases 
generated by the BL iteration. 
The indices $l$ and $m$ represent the site position of the resulting one-dimensional model and correspond to the ones 
in the symmetry adapted fermionic operator $\gamma_{l,m,\sigma}$ used in Eq.~(\ref{eq:hamq1dsym}). 
}
\label{fig:symbase}
\end{center}
\end{figure}

In the presence of the reflection or ${\rm C}_2$ rotation point group symmetry at the center of two impurity sites, 
the Q1D model $\mathcal{H}^{\rm Q1D}_{\rm WM}$ in Eq.~(\ref{eq:hamq1d}) can be further simplified by introducing 
symmetric and antisymmetric bases, 
\begin{eqnarray}
\gamma_{1, 1, \sigma} = (a_{1, 1, \sigma} + a_{1, 2, \sigma})/\sqrt{2} = (c_{{\bf r}_1,\sigma}+c_{{\bf r}_2,\sigma})/\sqrt{2}, \nonumber \\
\gamma_{1, 2, \sigma} = (a_{1, 1, \sigma} - a_{1, 2, \sigma})/\sqrt{2} = (c_{{\bf r}_1,\sigma}-c_{{\bf r}_2,\sigma})/\sqrt{2}, \label{eq:sasop}
\end{eqnarray}
as the initial BL bases for the BL iteration. It is then readily shown that the two-impurity Wolff model $\mathcal H_{\rm WM}$ in 
Eq.~(\ref{eq:ham}) can be mapped onto the following Q1D model: 
\begin{eqnarray}
\mathcal{\tilde H}^{\rm Q1D}_{\rm WM} &=& \sum_{m=1}^2\frac{U_{{\bf r}_m}}{4}
                   \left[ (\gamma_{1,1,\uparrow}^{\dagger}+(-1)^{m+1}\gamma_{1,2,\uparrow}^{\dagger})\right. \nonumber \\
&{}& \quad \quad \quad \left. \times (\gamma_{1,1,\uparrow}+(-1)^{m+1}\gamma_{1,2,\uparrow}) - 1 \right]\nonumber \\
&{}& \quad \quad \quad \times \left[ (\gamma_{1,1,\downarrow}^{\dagger}+(-1)^{m+1}\gamma_{1,2,\downarrow}^{\dagger}) \right. \nonumber \\
&{}& \quad \quad \quad \times \left. (\gamma_{1,1,\downarrow}+(-1)^{m+1}\gamma_{1,2,\downarrow})-1 \right] \nonumber \\
&+& \sum_{l=1}^L \sum_{m=1}^2\sum_\sigma \tilde\epsilon_{m}^l \gamma_{l,m,\sigma}^{\dagger} \gamma_{l,m,\sigma} \nonumber \\
&+& \sum_{l=1}^{L-1} \sum_{m=1}^2\sum_\sigma \tilde t_{m}^l \left( \gamma_{l,m,\sigma}^{\dagger} \gamma_{l+1,m,\sigma} + {\rm h.c.}\right). \label{eq:hamq1dsym}
\end{eqnarray}
Here, $\gamma_{l,m,\sigma}$ is the $l$th BL bases generated by the BL recursive technique with the initial BL bases 
$\gamma_{1,m,\sigma}$ given in Eq.~(\ref{eq:sasop}). 
It should be noticed that, in contrast to the previous Q1D ladder model $\mathcal{H}^{\rm Q1D}_{\rm WM}$ in Eq.~(\ref{eq:hamq1d}), 
the resulting Q1D model $\mathcal{\tilde H}^{\rm Q1D}_{\rm WM}$ is now completely decoupled [see Fig.~\ref{fig:symbase}(a)], 
owing to the symmetry adapted BL bases, except for the ``initial'' sites ($l=1$), i.e., the interacting impurity sites. 
This form is particularly useful for the DMRG calculations because this Q1D model is regarded as a pure one-dimensional 
chain model, as schematically shown in Fig.~\ref{fig:symbase}(b). We should also note that, because of this choice of the initial 
BL bases in Eq.~(\ref{eq:sasop}), the two-body Coulomb interaction terms in $\mathcal{\tilde H}^{\rm Q1D}_{\rm WM}$ contain intersite 
interactions between the impurity sites, although in the original representation of the Wolff model $\mathcal H_{\rm WM}$ 
the two-body interaction terms are local. However, this slight complexity does not cause any difficulty in applying the DMRG method to 
$\mathcal{\tilde H}^{\rm Q1D}_{\rm WM}$.

Three remarks are in order. 
First, exactly the same pure one-dimensional Hamiltonian $\mathcal{\tilde H}^{\rm Q1D}_{\rm WM}$ given in 
Eq.~(\ref{eq:hamq1dsym}), including the two-body interaction part, can be 
constructed by using the standard Lanczos tridiagonalization procedure applied separately to each symmetric and antisymmetric 
basis given in Eq.~(\ref{eq:sasop}) as the initial Lanczos basis. Indeed, a similar idea using the standard Lanczos 
tridiagonalization procedure has been proposed for two-impurity models in the context of NRG~\cite{jones,silva}. 
Second, for the two-impurity Wolff model on the honeycomb lattice defined in Eqs.~(\ref{eq:ham})--(\ref{eq:hamimp}), 
the symmetric and antisymmetric BL bases in Eq.~(\ref{eq:sasop}) can always decouple the Q1D ladder model 
$\mathcal H_{\rm WM}^{\rm Q1D}$ to a pure 
one-dimensional model $\mathcal {\tilde H}_{\rm WM}^{\rm Q1D}$, regardless of the location of two impurity sites. 
Third, this simplification is made possible solely because of the symmetry of the one-body part of the original 
Hamiltonian $\mathcal H_{\rm WM}$. 
The similar simplification of the Q1D model using the symmetry adapted BL bases can be applied to more involved models, 
an example being discussed below in Sec.~\ref{sec:moam}.

\begin{figure*}[htbp]
\begin{center}
\includegraphics[width=\hsize]{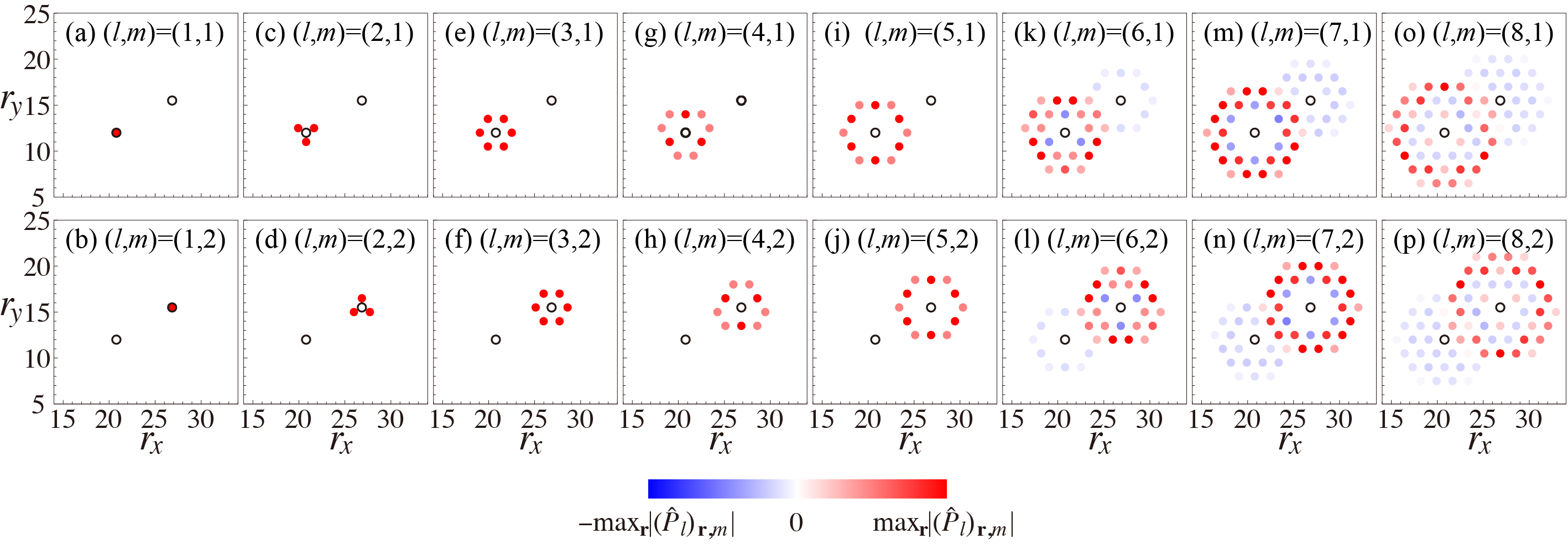}
\caption{(Color online) Intensity plots of $(\hat{P_l})_{{\bf r},m}$ at ${\bf r}=(r_x,r_y)$, i.e., the real-space distribution of 
the $m$th BL bases generated after the $l$th BL iteration, for the two-impurity Wolff model on the honeycomb lattice 
$\mathcal H_{\rm WM}$ in Eq.~(\ref{eq:ham}). 
The BL bases for different $l$ and $m$ (indicated in the figures) 
are generated starting with the initial BL bases $\hat P_1$ given in Eq.~(\ref{eq:blinit}), thus corresponding to $a^\dag_{l,m,\sigma}$ 
used in ${\mathcal H}_{\rm WM}^{\rm Q1D}$ [Eq.~(\ref{eq:hamq1d})]. 
The two impurity sites are located at ${\bf r}_1=(12\sqrt{3},12)$ and 
${\bf r}_2=(15.5\sqrt{3},15.5)$, indicated by open black circles. 
}
\label{fig:ripple}
\end{center}
\end{figure*}

Let us now discuss the physical meaning of the BL bases generated by the BL recursive technique  
for the two-impurity Wolff model $\mathcal H_{\rm WM}$ on the honeycomb lattice. 
Since $a^\dag_{l,m,\sigma}=\sum_{\bf r} c^\dag_{{\bf r},\sigma}( \hat P_l )_{{\bf r},m}$ [see Eq.~(\ref{eq:ut_a})], 
the $m$th BL bases generated after the $l$th BL iteration is represented by $(\hat P_l)_{{\bf r},m}$, 
where ${\bf r}$ is a two-dimensional vector on the honeycomb lattice. 
Figure~\ref{fig:ripple} shows ${\bf r}$ dependence of 
$(\hat{P_l})_{{\bf r},m}$ for $m=1$ and $2$ obtained with the initial BL bases 
$\hat P_1$ given in Eq.~(\ref{eq:blinit}). 
In the standard Lanczos tridiagonalization procedure with the initial Lanczos basis similar to $\hat P_1$, 
e.g., ${\bm e}_1$ in Eq.~(\ref{eq:blinit}), the Lanczos basis generated after the $l$th Lanczos iteration 
forms a $s$-wave ``ripple'' around the impurity site for any $l$ and the size of the ripple increases 
with $l$~\cite{krishna-murthy,borda,brusser}. 
Similarly, as shown in Figs.~\ref{fig:ripple}(a)--\ref{fig:ripple}(j), every BL iteration 
generates two orthogonal bases for $m=1$ and 2, and each basis is like a propagating ripple centered 
at each impurity site. However, these BL bases generated are no longer $s$-wave-like once they overlap. 
This is simply because these two bases must be orthogonal and thus 
they can not be $s$-wave-like once these two ripples overlap each other [see Figs.~\ref{fig:ripple}(k)-\ref{fig:ripple}(p)].

\begin{figure*}[htbp]
\begin{center}
\includegraphics[width=\hsize]{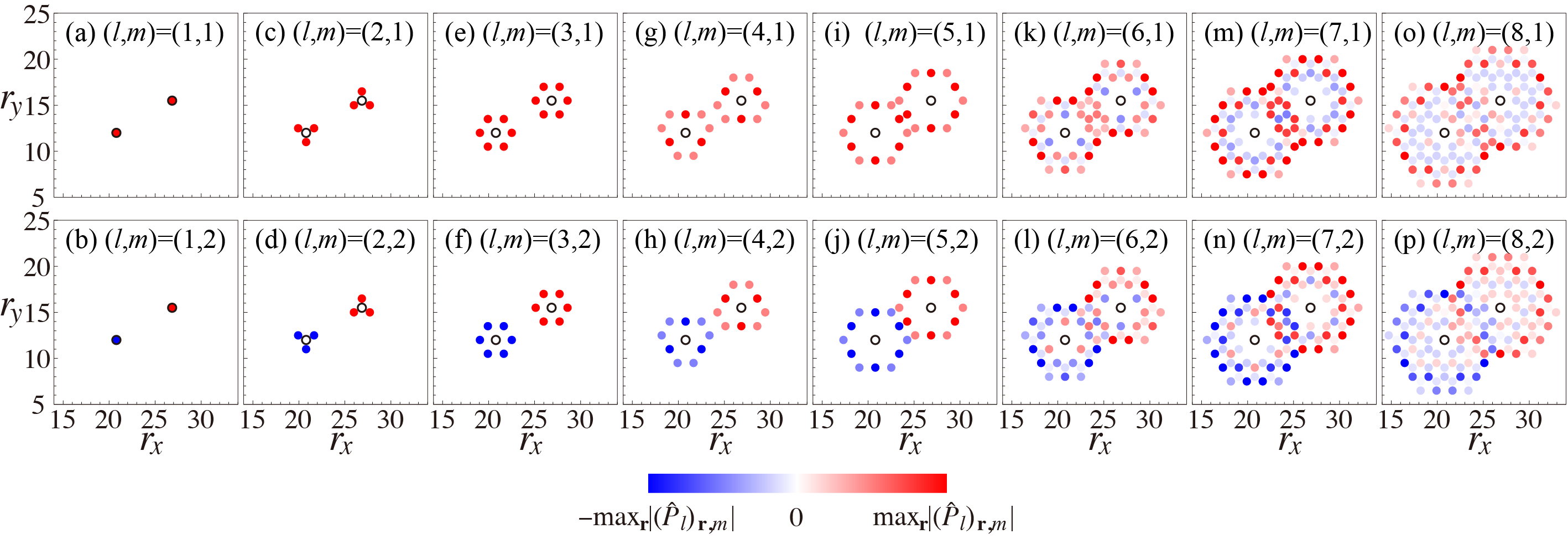}
\caption{(Color online) Intensity plots of $(\hat{P_l})_{{\bf r},m}$ at ${\bf r}=(r_x,r_y)$, i.e., the real-space distribution of 
the $m$th BL bases generated after the $l$th BL iteration, for the two-impurity Wolff model on the honeycomb lattice 
$\mathcal H_{\rm WM}$ in Eq.~(\ref{eq:ham}). 
The BL bases for different $l$ and $m$ (indicated in the figures) 
are generated starting with the symmetric and antisymmetric initial BL bases, $\gamma^\dag_{l, 1, \sigma}$ and 
$\gamma^\dag_{l, 2, \sigma}$, given in Eq.~(\ref{eq:sasop}), thus corresponding to $\gamma^\dag_{l, m, \sigma}$ 
used in $\tilde{\mathcal H}_{\rm WM}^{\rm Q1D}$ [Eq.~(\ref{eq:hamq1dsym})].  
The two impurity sites are located at ${\bf r}_1=(12\sqrt{3},12)$ and 
${\bf r}_2=(15.5\sqrt{3},15.5)$, indicated by open black circles. 
}
\label{fig:ripple_sym}
\end{center}
\end{figure*}

It is also interesting to see the ripples for $\gamma^\dag_{l,m,\sigma}$ generated after the $l$th BL iterations using the symmetric 
and antisymmetric initial BL bases given in Eq.~(\ref{eq:sasop}). 
In general, the off-diagonal terms in $\hat{E}_l$ as well as $\hat{T}_l$ are due to the interference between the two ripples for 
$m=1$ and $2$ once the two ripples overlap [see Figs.~\ref{fig:ripple}(k)-\ref{fig:ripple}(p)]. 
However, because the BL iteration respects the symmetry of Hamiltonian 
$\mathcal H_0^{\rm WM}=\mathcal H_c^{\rm WM} + \mathcal H_V^{\rm WM}$, the BL bases generated 
still preserve the symmetric and antisymmetric characteristics even for $l>1$ if the symmetric and antisymmetric initial BL bases 
are used. This can be clearly seen in Fig.~\ref{fig:ripple_sym}. 
Both before [Figs.~\ref{fig:ripple_sym}(a)-\ref{fig:ripple_sym}(j)] and after [Figs.~\ref{fig:ripple_sym}(k)-\ref{fig:ripple_sym}(p)] 
the two ripples overlap, they are clearly symmetric and antisymmetric with respect to 
${\rm C}_2$ rotation (or reflection) at the center of two impurity sites.  
Therefore the off-diagonal elements in $\hat{E}_l$ and $\hat{T}_l$ are zero when the symmetry adapted BL bases are appropriately used.

Indeed, $a_{l,m,\sigma}$ and $\gamma_{l,m,\sigma}$ relate to each other and the relation depends on the relative position 
of two impurity sites on the honeycomb lattice. 
When the two impurity sites are located on different sublattices of the honeycomb lattice, 
the parameters $\epsilon_{mm^{\prime}}^l $ and $ t_{mm^{\prime}}^l $ in Eq.~(\ref{eq:hamq1d}) satisfy 
\begin{eqnarray}
&{}& \epsilon_{11}^l=\epsilon_{22}^l, \quad \epsilon_{12}^l=\epsilon_{21}^l, \nonumber \\
&{}&  t_{11}^l= t_{22}^l, \quad  t_{12}^l= t_{21}^l=0.
\end{eqnarray}
Therefore, in this case, $\gamma_{l,m,\sigma}$ for any $l\,(>1)$ is related to $a_{l,m,\sigma}$ via the following simple relations: 
\begin{eqnarray}
&{}& \gamma_{l, 1, \sigma} = (a_{l, 1, \sigma}+ a_{l, 2, \sigma})/\sqrt{2}, \nonumber \\
&{}& \gamma_{l, 2, \sigma} = (a_{l, 1, \sigma}- a_{l, 2, \sigma})/\sqrt{2}.
\end{eqnarray}
On the other hands, when the two impurity sites are located on the same sublattices, 
the parameters in Eq.~(\ref{eq:hamq1d}) satisfy 
\begin{eqnarray}
&{}& \epsilon_{11}^l = \epsilon_{12}^l = \epsilon_{21}^l = \epsilon_{22}^l = 0. 
\end{eqnarray}
Therefore, $\gamma_{l,m,\sigma}$ are determined so as to diagonalize $2\times2$ matrix 
$ t_{mm^{\prime}}^l $ with respect to $m$ and $m'$ in Eq.~(\ref{eq:hamq1d}).

\begin{figure}[htbp]
\includegraphics[width=0.95\hsize]{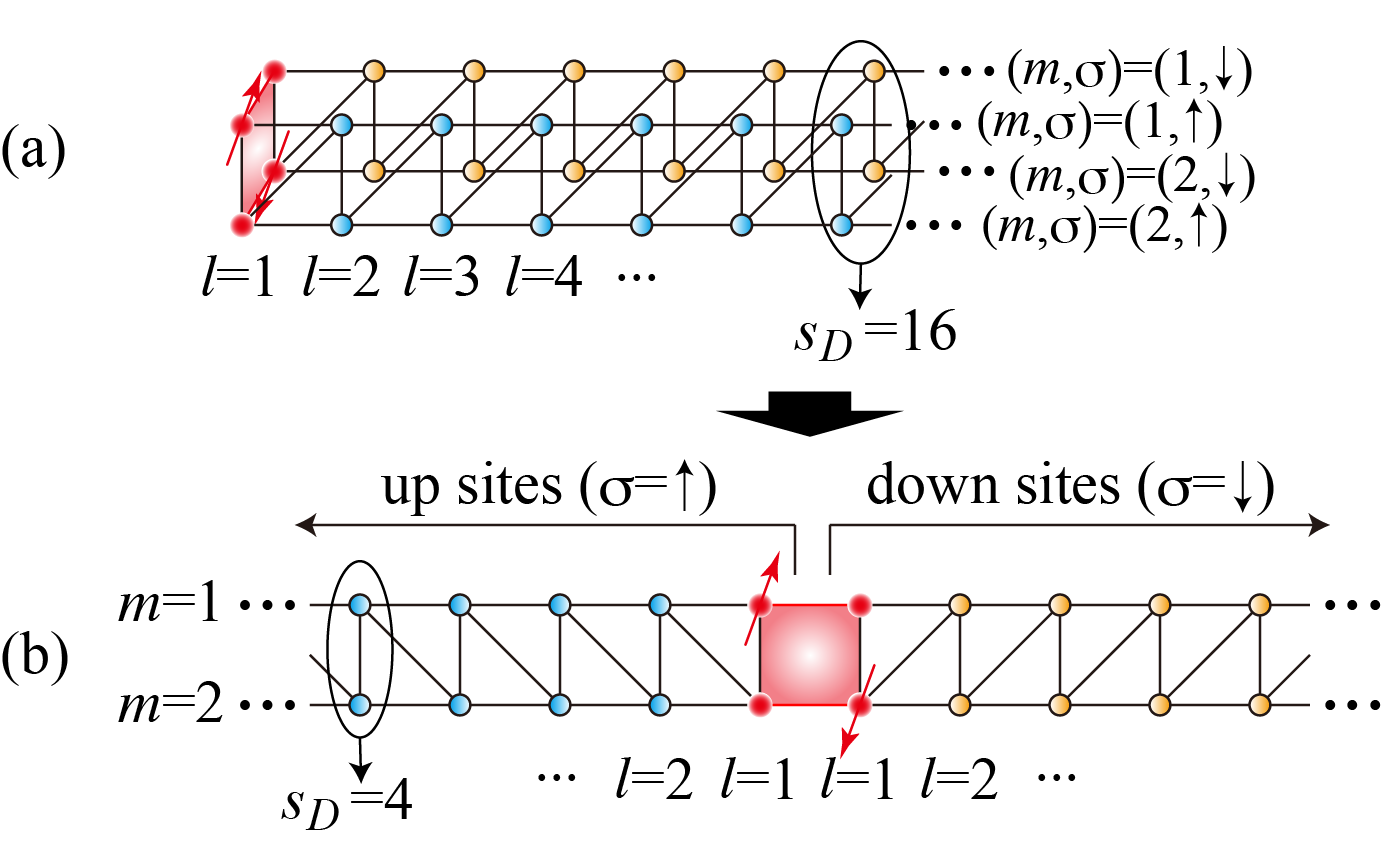}
\caption{(Color online) Schematic representation of a reduction scheme to save the computational cost 
in DMRG calculations by using spin degrees of freedom for a single-orbital two-impurity Anderson model 
${\mathcal H}_{\rm AIM}$ with $N_i=2$, $N_d=1$, and $M=2$. 
(a) The semi-infinite ladder model obtained by the BL recursive technique. 
Here the up and down electron sites (indicated 
by cyan and orange circles, respectively) are explicitly represented. 
The impurity sites are denoted by red spheres at the left edge. The red shaded plaquette at the left edge 
indicates where the two-body interaction part ${\mathcal H}_U$ is active at the impurity sites. 
The local degrees of freedom at each rung in this representation is $s_D=16$. 
(b) Using the fact that the one-body part of the Q1D model is decoupled for up and down electron sites (except for the impurity sites), 
the up electron part in (a) can be simply stretched to the left to form  
an infinite ladder model, which contains less local degrees of freedom at each rung, i.e, $s_D=4$. 
The indices $l$ and $m$, representing the site position of the ladder model along the leg and rung directions, 
respectively, correspond to the ones in the new fermionic operator $({\bm a}_{\sigma})_{lm}=a_{l,m,\sigma}$ used in 
Eq.~(\ref{eq:hamladt}). 
 }
\label{fig:smap}
\end{figure}

Finally, we note briefly another scheme which can be used to reduce the computational cost in DMRG calculations. 
This can be applied when the one-body part of the Hamiltonian is separated for up and down electrons, 
as in $\mathcal H_{\rm AIM}$ [Eq.~(\ref{eq:generalham})]~\cite{note3}. 
In this case, the one-body part of the Q1D model obtained by the BL recursive technique 
is also separated for up and down electrons [see $\mathcal H^{\rm Q1D}_{\rm AIM}$ in Eq.~(\ref{eq:hamlad})]. 
Therefore, the Q1D model is described by two decoupled semi-infinite Q1D Hamiltonians, 
one for up electron sites and the other for down electron sites, which connect to each other via two-body part of the Hamiltonian 
at the impurity sites, 
as schematically shown in Fig.~\ref{fig:smap}(a). By stretching the up electron part of the Q1D Hamiltonian to the left, 
we can finally obtain the infinite Q1D model, as shown in Fig.~\ref{fig:smap}(b). 

The total Hilbert space in DMRG calculations is proportional to $s^2_D m^2_D$, where $m_D$ is the 
number of density-matrix eigenstates kept in DMRG calculations and $s_D$ is the number of local states 
for added sites, i.e., the number of local states at each rung in the Q1D model. Therefore, in the case of 
two-impurity Wolff model $\mathcal H_{\rm WM}$, 
$s_D$ is reduced form 16 to 4 by using this reduction scheme for spin degrees of freedom. 
Although we can no longer use the fact that 
the total $S_z$ is a good quantum number to reduce the dimension of the Hilbert space, we find that this reduction scheme is still useful 
when there is no point group symmetry available to construct the symmetry adapted BL bases. 
We will use this reduction scheme in Sec.~\ref{sec:sscf} for systems where the symmetry adapted BL bases are not easily constructed.

\subsection{\label{sec:orbs}Calculations for spatially dependent quantities}

\begin{figure}[htbp]
\begin{center}
\includegraphics[width=65mm]{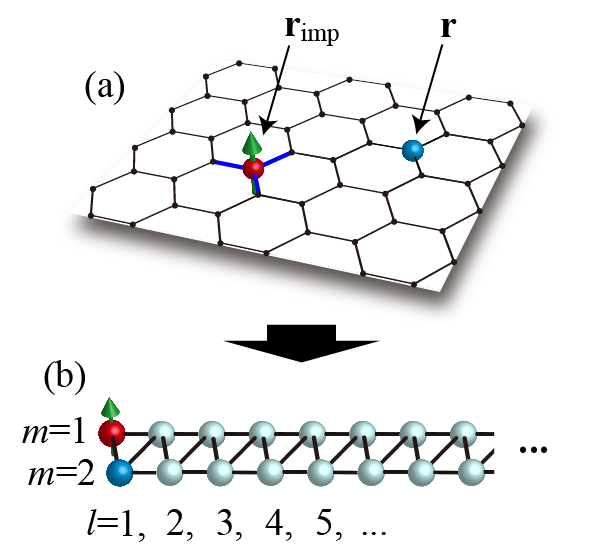}
\caption{(Color online) Schematic representation of the Q1D mapping for a single-impurity Wolff model on the 
honeycomb lattice to calculate correlation functions between the impurity site (denoted by red sphere with green arrow) 
at ${\bf r}_{\rm imp}$ and a conduction site (denoted by blue sphere) at ${\bf r}$. 
White spheres in (b) indicate the ripple states (i.e., BL bases) generated by the BL recursive technique with taking the impurity site 
${\bf r}_{\rm imp}$ and the conduction site ${\bf r}$ as the initial BL bases. 
The indices $l$ and $m$ in (b) correspond to the site position of the resulting semi-infinite ladder model along the leg and rung directions, 
respectively. 
The same Q1D mapping is used to calculate, e.g., local density of states at the conduction site ${\bf r}$. 
}
\label{fig:orbs}
\end{center}
\end{figure}

The BL-DMRG method allows us to 
calculate spatially dependent quantities in real space, such as correlation functions between any sites 
and local density of states at any conduction sites. For example, to calculate correlation functions 
between the impurity site ${\bf r}_{\rm imp}$ and the conduction site ${\bf r}$, 
we can simply take the impurity site(s) and the conduction site of interest as the initial BL bases. 
The resulting Q1D model constructed by the BL recursive technique contains explicitly 
the impurity site ${\bf r}_{\rm imp}$ as well as the conduction site ${\bf r}$, for which the correlation functions are readily evaluated 
using the DMRG method.
This scheme is explained schematically for a single-impurity Wolff model in Fig.~\ref{fig:orbs}.

Although a similar idea has been applied in the NRG method~\cite{borda}, 
the BL-DMRG method has several advantages over the NRG method in calculating spatially dependent quantities: 
(i) the BL-DMRG method can treat any conduction Hamiltonians in real space,  
(ii) the reduction scheme to save the computational cost is available 
for the BL-DMRG method by using the symmetry adapted BL bases if the one-body part of the Hamiltonian has 
an appropriate symmetry, and (iii) the reduction scheme using spin degrees of freedom can also be applied in the BL-DMRG method if 
the one-body part of the Hamiltonian is separated for up and down electrons.

Instead of constructing a different Q1D model for each conduction site of interest, as shown in Fig.~\ref{fig:orbs} (b), 
it is in principle possible to calculate physical quantities involving the conduction sites 
by using Eq.~(\ref{eq:ut_c}) directly for a single Q1D model constructed with the initial BL bases containing only 
the impurity sites. However, this approach suffers several problems. 
First of all, it is not necessarily true that the well-defined nonsingular $(N+M)\times(N+M)$ unitary matrix $\hat P$ in Eq.~(\ref{eq:pdef}) 
is always obtained by the BL iterations in Eq.~(\ref{eq:bliter}). This is simply because the BL bases generated by 
the BL recursive technique belong to a certain irreducible representation determined by the initial BL bases. 
The bases belonging to other representations are not generated because these bases are decoupled to the impurity sites. 
Second, the BL iterations are very often terminated with a finite number $L$ of iterations, specially when we consider 
the conduction sites in the thermodynamics limit $N\to\infty$. In this case, $\hat P$ is a rectangular $(N+M)\times(LM)$ matrix 
and thus the inverse of $\hat P$ can not be defined to describe the operator ${\bf c}_\sigma^\dag$ for the conduction sites 
using the BL bases operator ${\bf a}_\sigma^\dag$~\cite{note5}. 
In spite of all these difficulties, if we obtained the well-defined unitary 
matrix $\hat P$, we would then represent the physical quantities using the BL bases, e.g., 
\begin{eqnarray}
c_{n,\sigma}^{\dagger} c_{n',\sigma'} = \sum_{l,l^{\prime}} \sum_{m,m^{\prime}} 
(\hat{P}^{\dagger}_l)_{m,n} (\hat{P}_{l^{\prime}})_{n',m^{\prime}} 
a_{l,m,\sigma}^{\dagger} a_{l^{\prime},m^{\prime},\sigma'}.  
\end{eqnarray}
However, we would still have to carry out these matrix multiplications for all 
$a_{l,m,\sigma}^{\dagger} a_{l^{\prime},m^{\prime},\sigma}$, separately, 
which is computationally very demanding. 
On the other hand, any operator involving the conduction sites can be incorporated exactly in the Q1D model 
generated by the BL recursive technique if the conduction sites are included explicitly in the initial BL bases (see Fig.~\ref{fig:orbs}).

\subsection{\label{sec:moam}Multiorbital systems}

It is rather straightforward to extend the BL-DMRG method for a multiple-impurity Anderson model to 
a multiorbital single-impurity Anderson model. For completeness and for possible future applications, we shall here briefly 
describe the formulation of the BL-DMRG method for a multiorbital single-impurity Anderson model and discuss the 
symmetry of the BL bases.

\begin{figure}[htbp]
\begin{center}
\includegraphics[width=80mm]{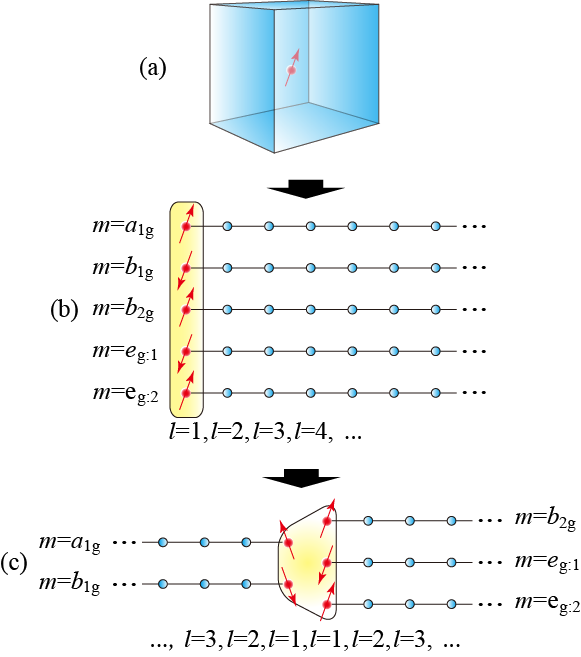}
\caption{(Color online) Schematic representation of the Q1D mapping for a five $d$-orbital single-impurity 
Anderson model. The BL recursive technique transforms the five $d$-orbital single-impurity Anderson model 
onto a semi-infinite five-leg ladder model. A blue cube represents the conduction sites in (a) and 
blue circles indicate the BL bases in (b) and (c). 
Red circles with arrows denote the impurity site with orbital $m$ (=$a_{1g}$, $b_{1g}$, $b_{2g}$, $e_{g:1}$, and $e_{g:2}$). 
The yellow shaded regions in (b) and (c) indicate where the two-body Coulomb interaction ${\mathcal H}_U^d$ is active at the impurity sites. 
The indices $l$ and $m$ in (b) and (c), representing the site position of the resulting Q1D model along the leg and rung directions, 
respectively, correspond to the ones in the new fermionic operator $({\bm a}_{\sigma})_{lm}=a_{l,m,\sigma}$ in 
Eq.~(\ref{a_op}) and used to describe the Q1D model in Eq.~(\ref{eq:hamlad}). 
}
\label{fig:d5map}
\end{center}
\end{figure}

As an example, we shall consider a five $d$-orbital single-impurity Anderson model. 
The Hamiltonian is given by Eq.~(\ref{eq:generalham}) with $N_i=1$ and $N_d=5$. 
Assuming that the impurity site is in a tetragonal environment with $D_{4h}$ point group symmetry, 
the five fold degenerate $d$ orbitals are reducible and contain the following irreducible representations: 
$a_{1g}$ ($d_{3z^2-r^2}$ orbital), $b_{1g}$ ($d_{x^2-y^2}$ orbital), $b_{2g}$ ($d_{xy}$ orbital), and 
($e_{g:1}$, $e_{g:2}$) [($d_{yz}$, $d_{zx}$) orbitals]. 
For the two-body part of the Hamiltonian, we can consider, e.g., the most complete interactions, 
\begin{eqnarray}
\mathcal{H}_U^{d} &=& U \sum_{m} n_{m,\uparrow} n_{m,\downarrow} + U^{\prime} \sum_{m < m^{\prime}} 
\sum_{\sigma} n_{m,\sigma} n_{m^{\prime},\bar{\sigma}} \nonumber \\
&{}& + (U^{\prime}- J) \sum_{m < m^{\prime}} n_{m,\sigma} n_{m^{\prime},\sigma}  \nonumber \\
&{}& - J \sum_{m \neq m^{\prime}} 
c_{m,\uparrow}^{\dagger} c_{m,\downarrow} c_{m^{\prime},\downarrow}^{\dagger} c_{m^{\prime},\uparrow} \nonumber \\
&{}& + J^{\prime} \sum_{m \neq m^{\prime}} c_{m,\uparrow}^{\dagger} c_{m,\downarrow}^{\dagger} 
c_{m^{\prime},\downarrow} c_{m^{\prime},\uparrow}, \label{eq:multiorbit}
\end{eqnarray}
where $U$, $U^{\prime}$, $J$, and $J^{\prime}$ are intra-orbital Coulomb interaction, 
interorbital Coulomb interaction, Hund's coupling, and pair-hopping, respectively. Here, 
$m=(a_{1g}, b_{1g}, b_{2g}, e_{g:1}, e_{g:2})$, 
$n_{m,\sigma} = d_{m,\sigma}^{\dagger} d_{m,\sigma}$ and $\bar{\sigma}$ indicates the opposite spin of $\sigma$. 
Applying the BL recursive technique, the five $d$-orbital single-impurity Anderson model is mapped onto a semi-infinite five-leg ladder model, 
as shown in Fig.~\ref{fig:d5map}. 

Let us now discuss the symmetries of the BL bases, i.e., ripple states, generated by the BL recursive technique. For simplicity, 
we further assume that the conduction bands coupled to the impurity site are formed by 
$s$ orbitals on the square lattice and the impurity site is embedded in one of the sites forming the square lattice. 
Then, the five $d$-orbital single-impurity Anderson model is describe by the following Hamiltonian: 
\begin{equation}
 \mathcal{H}_{d} = \mathcal{H}_c^s + \mathcal{H}_V^{sd} + \mathcal{H}_U^d, \label{eq:egmodel}
 \end{equation}
 where
\begin{equation}
\mathcal{H}_c^s = - t{\sum_{\left< {\bf r}, {\bf r}^{\prime}\right>}}^\prime\sum_\sigma 
 \left ( c_{{\bf r},{\sigma}}^{\dagger} c_{{\bf r}',\sigma} + {\rm h.c.} \right )
\end{equation}
and
\begin{eqnarray}
\mathcal{H}_V^{sd} &=& V_{1} \sum_{{\bf e}= \pm{\bf e}_x, \pm{\bf e}_y} \sum_\sigma
\left ( c_{{\bf r}_{\rm imp}+{\bf e},\sigma}^{\dagger} d_{a_{1g},\sigma} + {\rm h.c.} \right)\nonumber \\
&+& V_{2} \sum_{{\bf e}= \pm{\bf e}_x} \sum_\sigma \left ( c_{{\bf r}_{\rm imp}+{\bf e},\sigma}^{\dagger} d_{b_{1g},\sigma} + {\rm h.c.}\right ) \nonumber \\
& -& V_{2} \sum_{{\bf e}= \pm{\bf e}_y} \sum_\sigma \left( c_{{\bf r}_{\rm imp}+{\bf e},\sigma}^{\dagger} d_{b_{1g},\sigma} + {\rm h.c.} \right).
\label{eq:egmodelv}
\end{eqnarray}
Here, the sum in ${\mathcal H}_c^s$ runs over all nearest-neighbor sites, ${\bf r}$ and ${\bf r}^{\prime}$, 
excluding the impurity site ${\bf r}_{\rm imp}$. 
$\mathcal{H}_V^{sd}$ represents the hybridization between the impurity site and the surrounding nearest-neighbor conduction sites. 
${\bf e}_x$ and ${\bf e}_y$ are the lattice unit vectors along $x$- and $y$-directions on the square lattice, respectively. 
The symmetry of the $d$ orbitals is reflected with the sign of the hybridization parameters in ${\mathcal H}_V^{sd}$ and also causes 
zero hybridization between $d_{xy}$, $d_{yz}$, and $d_{zx}$ orbitals and $s$ orbital. 
Moreover, here we simply ignore the one-body term at the impurity site.

\begin{figure}[htbp]
\begin{center}
\includegraphics[width=80mm]{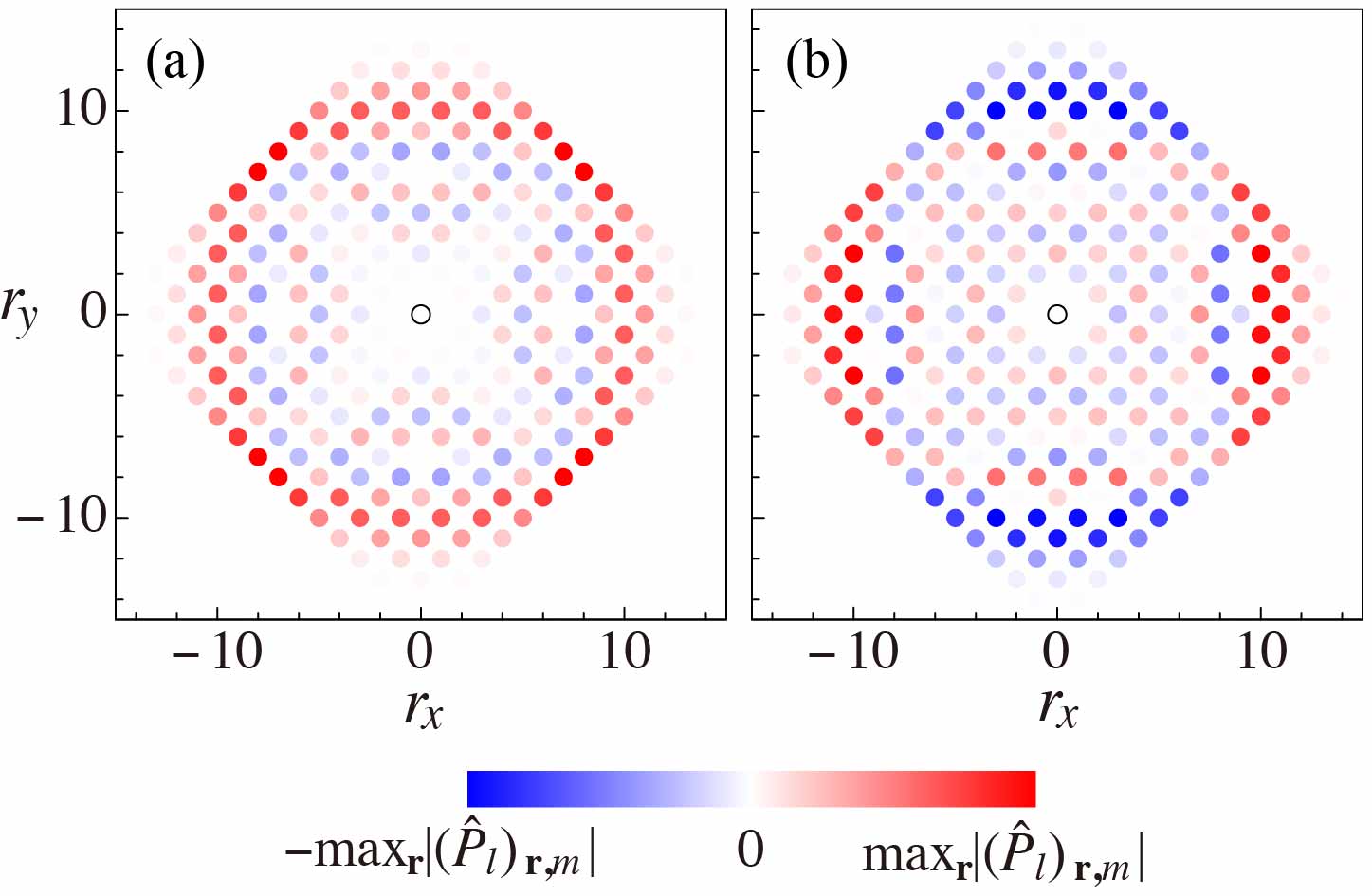}
\caption{(Color online) Intensity plots of ${\bf r}=(r_x,r_y)$ dependence of $(\hat{P}_l)_{{\bf r},m}$, i.e., 
the real-space distribution of the $m$-th BL bases generated after the $l$-th BL iteration, for the $d$-orbital single-impurity Anderson model 
on the square lattice ${\mathcal H}_d$ in Eq.~(\ref{eq:egmodel}). 
The BL bases for (a) $m = a_{1g}$ and (b) $m=b_{1g}$ 
are generated with taking the $d$ orbitals as the initial BL bases. 
Here, the results for $l = 15$ are shown. 
The impurity site is located at ${\bf r} = (0,0)$, indicated by open black circle. 
 }
\label{fig:dorbit}
\end{center}
\end{figure}

Applying the BL recursive technique with taking the $d$ orbitals as the initial BL bases, we can generate the BL bases which belong to the 
same irreducible representation with the initial BL bases. 
This can be best seen in ${\bf r}$ dependence of $(\hat{P}_l)_{{\bf r},m}$ [Eq.~(\ref{eq:ut_a})], 
i.e., the $m$th BL bases generated after the $l$th BL iteration. 
A typical example is shown in Fig.~\ref{fig:dorbit}. 
Although in this case only $d_{3z^2-r^2}$ and $d_{x^2-y^2}$ orbitals hybridize with the conduction $s$ orbitals, Fig.~\ref{fig:dorbit} 
clearly demonstrates that the each symmetry of the initial BL bases is preserved even after the BL iterations are executed. 
Because of the different irreducible representations, there is no matrix element in the resulting Q1D model between the BL bases 
$a_{l,m,\sigma}$ with different $m$'s for $l>1$ [see Fig.~\ref{fig:d5map}(b)].

Even when the conduction bands formed by $p$ orbitals are considered, the same conclusion is reached as long as 
the symmetry is respected correctly in the Hamiltonian. 
In this case, one can show that the five $d$ orbitals are all coupled to the conduction sites with finite hybridization, and 
the BL bases generated by the BL iterations preserve the same irreducible representation of the five $d$ orbitals when they 
are used for the initial BL bases. 
The resulting Q1D model is a semi-infinite five-leg ladder model, where different legs belong to different irreducible 
representations and thus the legs are 
decoupled to each other except for the impurity site, as shown in Fig.~\ref{fig:d5map}(b). 
More generally, in many cases, a multiorbital single-impurity Anderson model possesses a specific point group symmetry, 
and therefore the corresponding Q1D model is decoupled according to the irreducible representation of the bases~\cite{note4}.

Let us finally discuss the reduction scheme for the multiorbital systems to save the computational cost. 
First, it is trivial to apply the reduction scheme using the spin degrees of freedom (see Fig.~\ref{fig:smap}). 
Second, as shown above, a Q1D model mapped from a multiorbital single-impurity Anderson model is 
a semi-infinite ladder model with decoupled chains, except for the impurity site, because there is no matrix element between 
the bases with different irreducible representations. This can be used to reduce the computational cost by 
e.g., putting two of the five semi-infinite legs on the left and the other three on the right, ending up with an infinite ladder model 
with less number of legs, as shown in Fig.~\ref{fig:d5map}(c). 
Although this scheme introduces an imbalance of the Hilbert space between the left and right sides of the system 
when the impurity contains an odd number of orbitals, we still find this scheme to be very effective to save the computational cost.

The BL-DMRG method introduced here can be readily extended to any multiorbital multiple-impurity Anderson 
models. Therefore, we expect that the BL-DMRG method is efficiently applied as an impurity solver of DMFT 
for multiorbital Hubbard models~\cite{modmft} and for realistic electronic structure calculations of correlated materials~\cite{DFT+DMFT}. 
With straightforward extension, the BL-DMRG method is applied also to Kondo impurity models where localized spins are coupled to 
conduction sites.

\section{\label{sec:application}Application: Magnetic Impurity Problems in Graphene}

In this section, using the BL-DMRG method introduced in Sec.~\ref{sec:method}, 
we shall study three different single-impurity Anderson models for graphene with a single structural defect and 
with a single adatom. We first 
introduce the models in Sec.~\ref{sec:model} and explain briefly the numerical details in Sec.~\ref{sec:numerical}, 
followed by the numerical results for the local magnetic 
susceptibility in Sec.~\ref{sec:dssf}, the local electronic density of states in Sec.~\ref{sec:ldos}, and the spin-spin 
correlation functions between the impurity site and the conduction sites in Sec.~\ref{sec:sscf}.

\subsection{\label{sec:model}Models}

We study three different single-impurity Anderson models in this section. 
The Hamiltonians ${\mathcal H}_\Gamma$ of these three models ($\Gamma={\rm I},\,{\rm II}$, and III) are given as 
\begin{equation}
\mathcal{H}_\Gamma = \mathcal{H}_t + \mathcal{H}_V + \mathcal{H}_U,
\end{equation}
where 
\begin{eqnarray}
\mathcal{H}_t &=& - t {\sum_{\left< {\bf r}, {\bf r}^{\prime} \right>}} \sum_{\sigma } 
\left( c_{{\bf r},\sigma}^{\dagger} c_{{\bf r}^{\prime},\sigma} + {\rm h.c.} \right), \\
\mathcal{H}_V &=& V \sum_{{\bf r}\in \mathcal{S}} \sum_{\sigma } \left( 
c_{{\bf r},\sigma}^{\dagger} c_{{\bf r}_{\rm imp},\sigma} + {\rm h.c.} \right), 
\end{eqnarray}
and 
\begin{equation}
\mathcal{H}_U =
U ( n_{{\bf r}_{\rm imp},\uparrow} - 1/2 ) ( n_{{\bf r}_{\rm imp},\downarrow} - 1/2 ). 
\label{eq:gmu}
\end{equation}
Here, $c_{{\bf r},\sigma}^\dag$ ($c_{{\bf r},\sigma}$) is the electron creation (annihilation) operator at site ${\bf r}$ and 
spin $\sigma(=\uparrow,\downarrow)$. 
${\mathcal H}_t$ describes the conduction sites with the nearest-neighbor hopping $t$ and thus the sum for 
$\left< {\bf r}, {\bf r}^{\prime} \right>$ runs over all nearest-neighbor pairs of conduction sites at 
${\bf r}$ and ${\bf r}^{\prime}$ on the honeycomb lattice. 
${\mathcal H}_V$ describes the hybridization between the impurity site at ${\bf r}_{\rm imp}$ and the conduction site at ${\bf r}$ 
where the sum over ${\bf r}\in \mathcal{S}$ is taken the conduction sites connected to the impurity site through $V$. 
${\mathcal H}_U$ describes the impurity site with the on-site interaction $U$ and 
$n_{{\bf r}_{\rm imp},\sigma}=c_{{\bf r}_{\rm imp},\sigma}^\dag c_{{\bf r}_{\rm imp},\sigma}$. The models described by 
${\mathcal H}_\Gamma$ correspond to a special case of the general Anderson impurity model ${\mathcal H}_{\rm AIM}$ 
with $N_i=N_d=M=1$ in Eq.~(\ref{eq:generalham}).

\begin{figure}[htbp]
\begin{center}
\includegraphics[width=85mm]{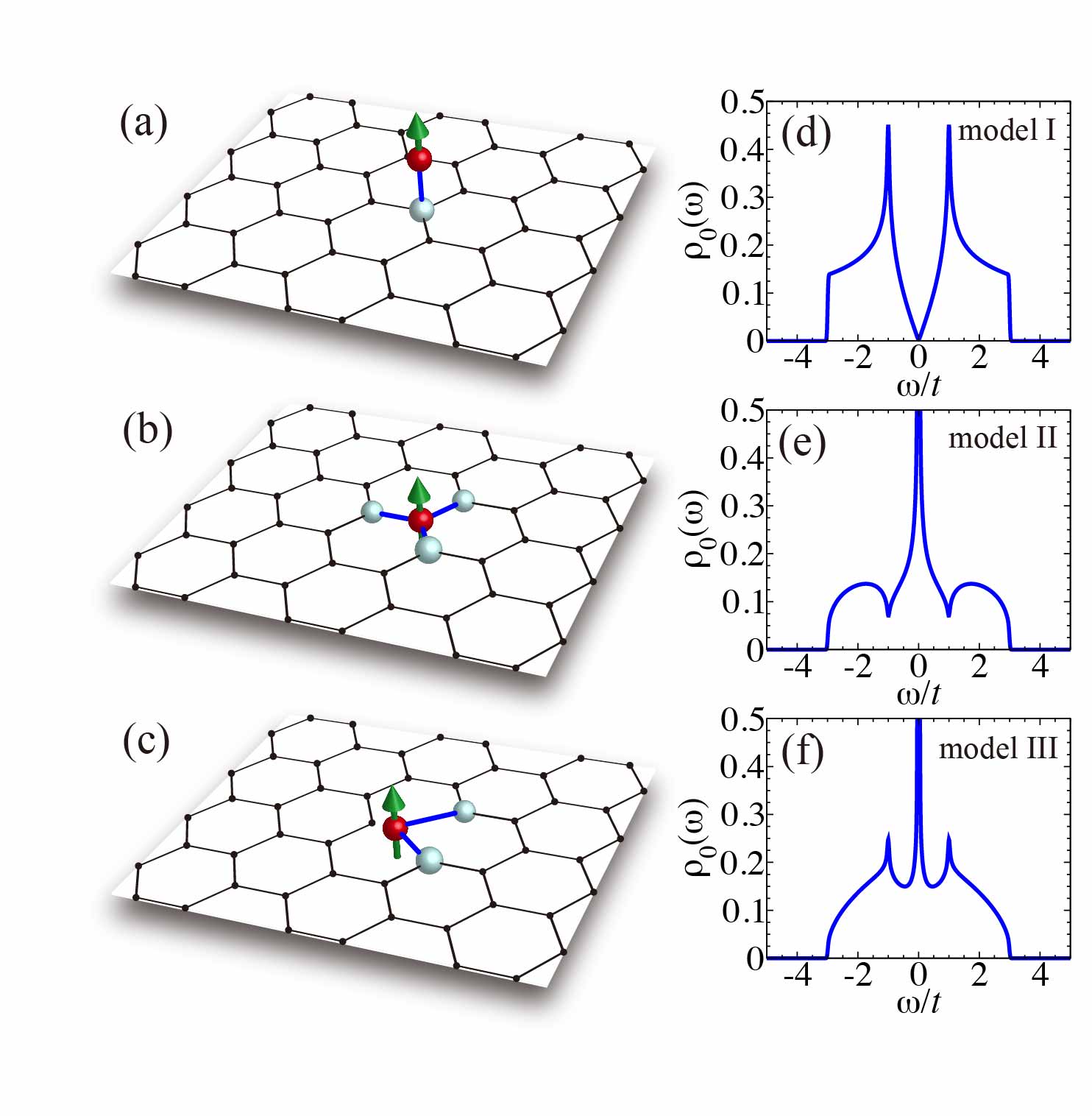}
\caption{(Color online) (a)--(c): Schematic representation of (a) a single adatom on the honeycomb lattice (model I), 
(b) a substitutional impurity in the honeycomb lattice (model II), and (c) an effective model for a single structural defect (vacancy) 
in graphene (model III). 
A red sphere with a green arrow indicates the impurity site. 
The conduction sites connected to the impurity site through the hybridization $V$ (denoted by bold blue lines) are 
represented by cyan spheres and other conduction sites are indicated by black dots. 
The hopping $t$ is finite only between the nearest-neighbor conduction sites, indicated by thin black lines. 
(d)--(f): Local density of states $\rho_0(\omega)$ per spin for $\mathcal H_t$ projected onto the second Lanczos basis 
with $l=2$ in Eq.~(\ref{eq:bliter}), 
i.e., the conduction sites connected to the impurity site through $V$, as indicated by cyan spheres in (a)--(c). 
Three models are indicated in figures (d)--(f). 
}
\label{fig:gmodel}
\end{center}
\end{figure}

These three models are different in the location of the impurity site 
and the way how the impurity site hybridizes with the conduction sites. 
The first model, model I, is for a single impurity absorbed (i.e., a single adatom) on the honeycomb lattice, 
as depicted in Fig.~\ref{fig:gmodel}(a). The impurity site is located on top of one of the conduction 
sites in the honeycomb lattice and hybridizes with only this conduction site. 
The second model, model II, represents a substitutional impurity in the honeycomb lattice, i.e., 
a single-impurity Wolff model, as depicted in Fig.~\ref{fig:gmodel}(b). One of the conduction sites in the honeycomb lattice 
is replaced by the impurity site, which hybridizes with the three nearest neighboring conduction sites. 
The third model, model III, represents an effective model for a single structural defect in graphene 
[see Fig.~\ref{fig:gmodel}(c)]. In this model, the impurity site is composed of a localized $sp^2$ dangling orbital, 
which hybridizes with the two neighboring sites, as indicated in Fig.~\ref{fig:gmodel}(c). Model III is obtained from model II 
with deleting one of the hybridizing bonds between the impurity site and the conduction sites in Model II.

\begin{figure}[htbp]
\begin{center}
\includegraphics[width=0.9\hsize]{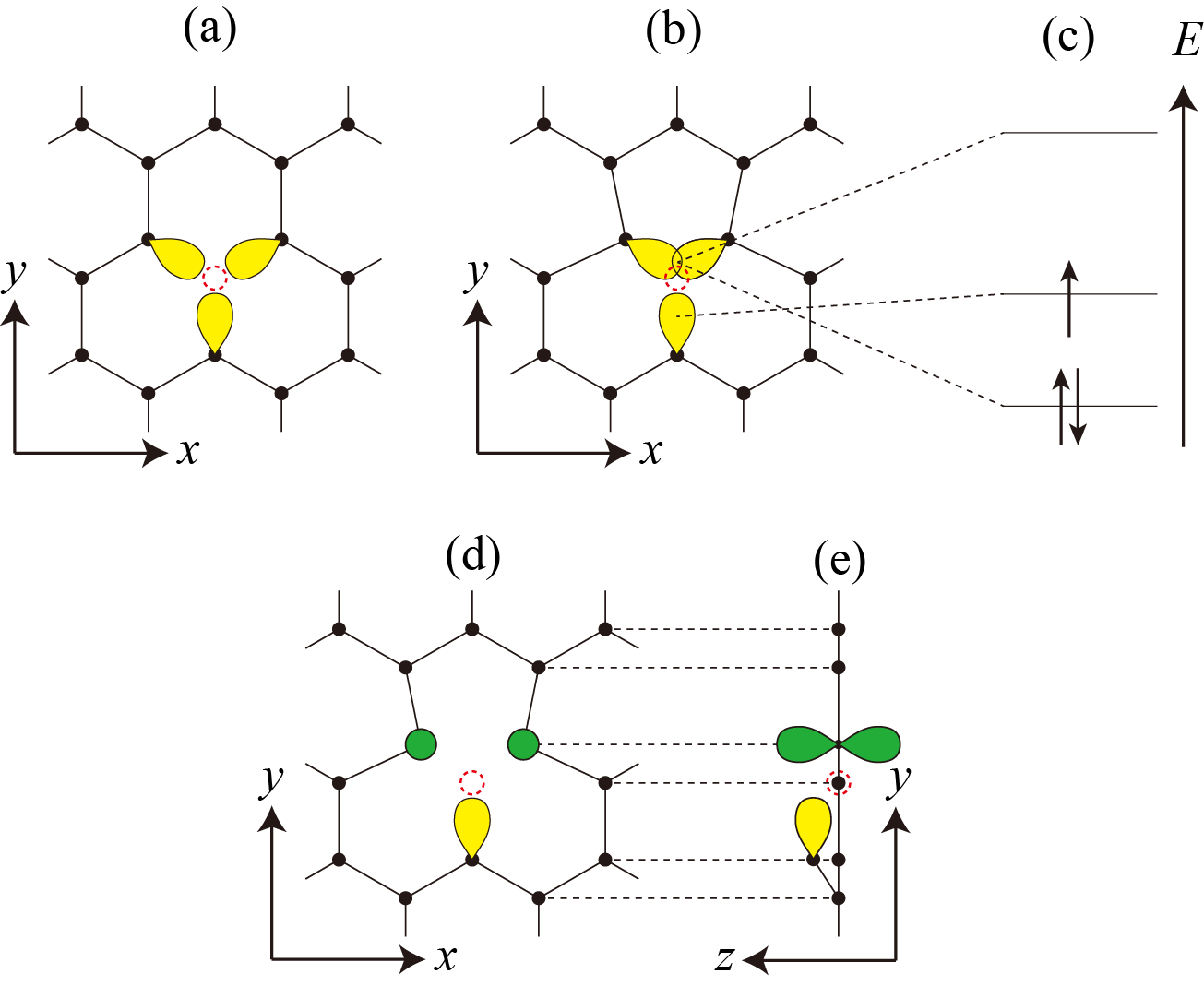}
\caption{(Color online) Schematic representation of local orbitals of carbon atoms (black dots) around the vacancy 
(red dashed circles) and a local molecular orbital energy diagram for model III~\cite{kanao}. 
(a) $sp^2$ dangling orbitals (yellow leaves) of the three carbon atoms surrounding the vacancy without structural distortion. 
(b) Same as (a) but with structural distortion reported by first-principles band structure 
calculations~\cite{barbary,yazyev,amara,palacios}. Two of the three carbon atoms surrounding the vacancy are closer to each other. 
(c) The resulting local molecular orbital energy diagram for (b). Without distortion and hybridization, 
the three dangling orbitals are degenerate. 
The lowest and highest levels correspond to the bonding and antibonding states, respectively, composed mostly of the dangling orbitals 
of the two carbon atoms closer to each other. The second lowest level corresponds to the nonbonding state 
composed mostly of the remaining dangling orbital. Since there are three electrons (arrows) in these dangling orbitals, 
the second lowest level is half-filled. 
(d) The half-filled dangling orbital (yellow leaf) and $p_z$ orbitals (green circles) of the other two neighboring carbon atoms 
surrounding the vacancy. 
Without additional distortion, the half-filled dangling orbital does not hybridize with other orbitals. 
(e) Same as (d) but the view from the in-plane axis of graphene. 
The green leaves indicate the $p_z$ orbitals. 
According to first-principles band structure calculations, the additional out-of-plane distortion takes place 
in the presence of vacancy~\cite{barbary,yazyev,amara,palacios}, which induces nonzero hybridization between the half-filled 
dangling orbital and the $p_z$ orbitals of the other two neighboring carbon atoms. 
}
\label{fig:model3}
\end{center}
\end{figure}

Model III deserves more explanation. In the presence of a single structural defect (i.e., vacancy) in graphene, three 
dangling orbitals appear around the defect, which are formed by $sp^2$ orbitals of three carbon atoms surrounding the defect, 
each carbon atom contributing a single $sp^2$ orbital, and 
are pointing towards the defect [see Fig.~\ref{fig:model3}(a)]. 
Without additional structural distortion and hybridization, these three dangling orbitals are degenerate. 
However, according to first-principles band-structure calculations~\cite{barbary,yazyev,amara,palacios}, 
because of the additional structural distortion around the defect, 
these three fold degenerate dangling orbitals are split into three nondegenerate levels, 
as shown in Figs.~\ref{fig:model3}(b) and \ref{fig:model3}(c). 
As a result, two of the three unpaired electrons in the $sp^2$ dangling orbitals occupy the lowest nondegenerate level 
and the remaining electron occupies the second lowest level, forming the localized state located mostly at one of the 
nearest neighboring carbon atoms around the defect~\cite{barbary,yazyev,amara,palacios}. 
Therefore, we can ignore the paired electrons occupying the lowest level and consider only the half-filled second lowest level 
as an impurity site. Note that the second lowest level is mainly composed of the $sp^2$ dangling orbital 
which points towards the defect and thus hybridizes mostly with the $p_z$ orbitals of the other two neighboring carbon atoms 
surrounding the defect, 
due to the additional out-of-plane distortion, as depicted in Figs.~\ref{fig:model3}(d) and \ref{fig:model3}(e), 
but not with the $p_z$ orbital at the same carbon atom because it is symmetrically forbidden~\cite{kanao}. 
Therefore, in model III, the $p_z$ orbital is also present at the same site where the impurity exists, although there is no direct hybridization 
between these two orbitals [see Fig.~\ref{fig:gmodel}(c)]. 
As mentioned above, the difference between model II and model III is the number of conduction sites which hybridize with 
the impurity site. 

As explained in details in Appendix~\ref{app:hybfunc}, the impurity properties of Anderson impurity models are determined solely 
by the hybridization function $\Delta (\omega)$~\cite{bulla2,bulla1}. 
The hybridization function for models I--III is expressed as
\begin{eqnarray}
\Delta (\omega) = \pi |T_1|^2 \rho_0 (\omega), \label{eq:hfunc}
\end{eqnarray}
where $T_1$ is the matrix element of $\mathcal H_t+\mathcal H_V$ between the first Lanczos basis, i.e., the impurity site, 
and the second Lanczos basis (see Sec.~\ref{sec:q1dmap} and Appendix~\ref{app:hybfunc}). As described in Appendix~\ref{app:hybfunc}, 
we can readily show that $T_1=V$ for model I, $T_2=\sqrt{3}V$ for model II, and 
$T_1=\sqrt{2}V$ for model III. 
$\rho_0(\omega)$ in Eq.~(\ref{eq:hfunc}) is 
the local density of states per spin for ${\mathcal H}_t$ projected onto the second Lanczos basis and is evaluated as 
\begin{eqnarray}
\rho_0 (\omega) = \frac{1}{N} \sum_{k=1}^N \left( \frac{1}{\sqrt{N_{\mathcal{S}}}}
\sum_{ {\bf r} \in \mathcal{S}} u_{{\bf r},c}^{(k)} \right)^2 \delta(\omega - \epsilon_{c,k}), 
\label{rho_0}
\end{eqnarray}
where $u_{{\bf r},c}^{(k)}$ is the $k$-th eigenstate of ${\mathcal H}_t$ at site ${\bf r}$ with its eigenvalue $\epsilon_{c,k}$. 
The sum over ${\bf r} \in \mathcal{S}$ in Eq.~(\ref{rho_0}) is taken for the conduction sites connected to the impurity site through $V$, 
as indicated by cyan spheres in Figs.~\ref{fig:gmodel}(a)--~\ref{fig:gmodel}(c), and $N_{\mathcal{S}}$ is the number of these sites. 
Since the hybridization function $\Delta (\omega)$ is proportional to the local density of state $\rho_0 (\omega)$, 
we can capture the fundamental difference among the three models simply by comparing $\rho_0 (\omega)$. 

As shown in Fig.~\ref{fig:gmodel}(d), $\rho_0 (\omega)$ for model I is exactly the same as  
the local density of states for the pure honeycomb lattice model. 
Therefore, model I is equivalent to the so-called pseudogap Kondo problem~\cite{bulla2,fritz1,vojta1,fritz2,ingersent}. 
The pseudogap Kondo problem has been studied both analytically and 
numerically based on the low-energy calculations~\cite{withoff,bulla2,gonzalez-baxton,ingersent,fritz2,vojta1,fritz1}.
The previous studies have found that the ground state is always in the local magnetic moment phase and 
hence no Kondo screening occurs as long as the system is the particle-hole symmetric. 
As will be shown below, our numerical calculations also find 
that the Kondo screening is absent for model I when the particle-hole symmetry is preserved at half filling.

In the case of models II and III, $\rho_0 (\omega)$ has a singularity at the Fermi level ($\omega = 0$), 
as shown in Figs.~\ref{fig:gmodel}(e) and \ref{fig:gmodel}(f). The appearance of the zero energy singularity is understood as follows.  
Recall first that $\rho_0 (\omega)$ is the local density of states for ${\mathcal H}_t$ projected onto the conduction sites next to 
the impurity site connected through $V$ in the honeycomb lattice. 
Therefore, assuming that these conduction sites belong to $B$ sublattice, 
the number $N_A$ of the conduction sites on $A$ sublattice  
is smaller by one than the number $N_B$ of the conduction sites on $B$ sublattice, i.e., $N_A=N_B-1$, where the total number $N$ of 
the conduction sites is $N_A+N_B$. 
Consequently, a single zero energy state is induced when there is no hopping between the same sublattices 
because the rank of $N\times N$ matrix for ${\mathcal H}_t$ is 
$N-1$. The zero energy state is localized mostly around the impurity site and the amplitude of the wave function of this state is 
finite only on $B$ sublattice. 
This zero energy state causes logarithmically diverging behavior in $\rho_0 (\omega)$ at 
$\omega =0$~\cite{vmpereira,peres}. 
We thus expect that the impurity properties for models II and III would be similar but different qualitatively from the one for model I.

\subsection{\label{sec:numerical}Numerical details}

As already indicated in Eq.~(\ref{eq:gmu}), in this paper, we consider only the particle-hole symmetric case at half filling. 
Therefore, the local electron density is always one, including at the impurity site, irrespectively of $U$ and $V$ values. 

To avoid unnecessary finite size effect~\cite{sorensen},
we always terminate the BL iteration at an even number $L$ of iterations when the Q1D model is constructed. 
Therefore, the resulting Q1D model has the even number $L$ of sites along the leg direction. 
For the calculations of physical quantities depending only on the impurity site, the resulting Q1D model is a pure one-dimensional 
chain and we consider $L$ up to $200$ with keeping 
$m_D \sim 12L$ density-matrix eigenstates in the DMRG calculations. 
For the calculations of physical quantities involving the conduction site, i.e., the spin-spin correlation functions between the impurity 
site and the conduction sites, the resulting Q1D model is a two-leg ladder model and we consider $L$ up to $240$ with keeping 
$m_D \sim 16L$ density-matrix eigenstates. The discarded weights are typically of the order $10^{-8}$ and the error of 
the ground state energy is $\sim 10^{-4}t$. 
We should emphasize that the resulting Q1D model with hundreds of $L$ sites along the leg direction 
corresponds to the original system $\mathcal H_\Gamma$ with tens of thousands of conduction sites $N$ in two spatial dimensions.
For example, the Q1D model with $L=240$ represents the original model $\mathcal H_\Gamma$ with at least $N\sim180,000$. 

To calculate the dynamical quantities, we employ the correction vector method~\cite{kuhner,jeckelmann}. 
Although the dynamical quantities can be evaluated with other methods, e.g., by expanding spectral functions 
into a continued fraction~\cite{hallberg3,dargel}, using Chebyshev polynomials~\cite{holzner2}, or Fourier transforming the 
corresponding real-time dynamics~\cite{rgpereira}, the correction vector method is most promising for our purpose 
because it is a direct calculation of the dynamical quantity by including the Hilbert space for the excited states 
and thus there is no additional error caused, e.g., by the 
numerical integration or by terminating the finite number of polynomials.

\subsection{\label{sec:results} Results}

\subsubsection{\label{sec:dssf}Local magnetic susceptibility at the impurity site}

Let us first examine the magnetic properties. For this purpose, 
here we calculate the local magnetic susceptibility $\chi_i (\omega)$ at the impurity site defined as 
\begin{eqnarray}
\chi_{i} (\omega) = -\frac{1}{\pi} {\rm Im} \left< \psi_0 \right| 
S_{{\bf r}_{\rm imp}}^z (\omega + i\eta + \mathcal{H}_\Gamma - E_0)^{-1} S_{{\bf r}_{\rm imp}}^z
\left| \psi_0 \right>, \nonumber \\
\end{eqnarray}
where $S_{{\bf r}_{\rm imp}}^z=(n_{{\bf r}_{\rm imp}, \uparrow}-n_{{\bf r}_{\rm imp}, \downarrow})/2$ is 
the $z$-component of spin operator at the impurity site ${\bf r}_{\rm imp}$, $\left| \psi_0 \right>$ is the ground state 
of $\mathcal H_\Gamma$ with its energy $E_0$, and $\eta\,(>0)$ is a broadening factor (a real number). 

In the noninteracting limit with $U=0$, $\chi_i (\omega)$ can be obtained directly using the $k$-th eigenstate ${\bf u}^{(k)}$ 
with its eigenvalue $\varepsilon_{k}$ of the one-body part of 
$\mathcal H_\Gamma$ described either by the conduction site bases ${\bf c}_{\sigma}$ as in 
$\hat{H}_0$ in Eq.~(\ref{eq:onebodyham}) or by the Lanczos bases ${\bf a}_{\sigma}$ as in $\hat{H}_0^{\rm BL}$ 
in Eq.~(\ref{eq:modonebodyham}), i.e., 
\begin{eqnarray}
\chi_i^0 (\omega) = \frac{\eta}{2\pi} \sum_{k \in (\varepsilon_{k} < \mu)} 
\sum_{k^{\prime} \in (\varepsilon_{k^{\prime}} > \mu)} \frac{ \left| u_{{\bf r}_{\rm imp}}^{(k)} u_{{\bf r}_{\rm imp}}^{(k^{\prime})} \right|^2 }
{ (\omega - \varepsilon_{k^{\prime}}+\varepsilon_{k})^2 + \eta^2}, \nonumber \\
\label{eq:nonintchi}
\end{eqnarray}
where $u_{\bf r}^{(k)}$ is the site ${\bf r}$ component of ${\bf u}^{(k)}$ and $\mu$  
is the chemical potential. Since models I, II, and III are all particle-hole symmetric at half filling, 
the chemical potential is $\mu = 0$. It is very intriguing to find that 
$\chi^0_i(\omega)$ can be calculated more accurately, in a sense that it is closer to the one in the thermodynamic 
limit, by using $\hat{H}_0^{\rm BL}$ than $\hat{H}_0$, as long as the same matrix sizes of $\hat{H}_0^{\rm BL}$ and $\hat{H}_0$ 
are taken. This is simply because more important degrees of freedom around the impurity site are extracted 
in $\hat{H}_0^{\rm BL}$ already for relatively small $L$.

\begin{figure}[htbp]
\includegraphics[width=\hsize]{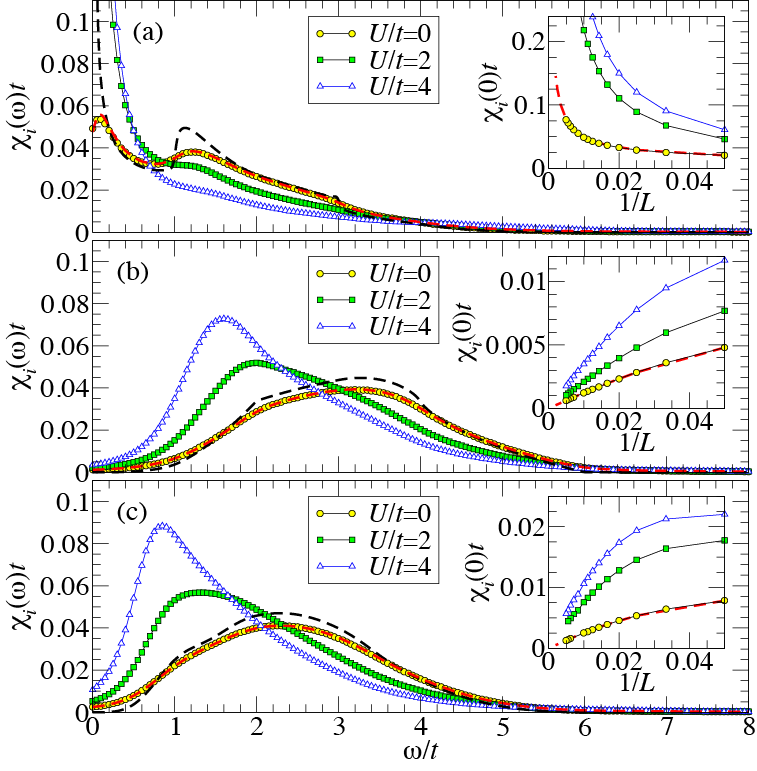}
\caption{(Color online) Local magnetic susceptibility $\chi_i (\omega)$ at the impurity site calculated 
for models (a) I, (b) II, and (c) III. 
The parameters used are $L=100$, $V=t$, and $\eta = 20 t/L$ for $U/t=0$ (circles), $2$ (squares), and $4$ (triangles). 
For comparison, $\chi_i^0 (\omega)$ for the noninteracting limit calculated using Eq.~(\ref{eq:nonintchi}) 
is also shown in red (black) dashed lines with $L=100$ $(1000)$ and $\eta = 20t/L$. 
Insets: $L$ dependence of $\chi_i(0)$ with keeping $\eta = 20/L$. For comparison, $\chi_i^0 (\omega)$ calculated using 
Eq.~(\ref{eq:nonintchi}) is also plotted by red dashed lines. 
}
\label{fig:msus}
\end{figure}

The results of $\chi^0_i(\omega)$ for the three different models in the noninteracting limit are shown in Fig.~\ref{fig:msus}. 
It is clearly observed in Fig.~\ref{fig:msus} that $\chi_i^0(\omega)$ diverges in the limit of $\omega\to0$ for model I while it converges to 
zero for models II and III. 
The different behavior of $\chi_i^0 (\omega)$ in the limit of $\omega\to0$ can be easily understood by recalling that 
$\chi^0_i(\omega)$ is proportional to the convolution of the local density of states, i.e., 
\begin{eqnarray}
\chi^0_i (\omega) \propto \int {\rm d}\omega \rho_i^0 (\omega' - \omega) \rho_i^0 (\omega^{\prime}) \Theta (\omega - \omega^{\prime}) \Theta (\omega^{\prime}), 
\end{eqnarray}
where $\Theta (\omega)$ is the Heaviside step function and $\rho_i^0 (\omega)$ is the local density of state at the impurity site 
[see Eq.~(\ref{eq:nonintrho})]. 
The diverging behavior of $\chi_i^0 (0)$ for model I is due to the presence of the zero energy state, 
which causes the zero energy peak at the Fermi level in the local density of state at the impurity site [see also in Fig.~\ref{fig:specv100}(a)]. 
In contrast, the local density of states at the impurity site for models II and III has the pseudogap structure at the Fermi level, i.e., 
$\rho_i^0 (\omega) \propto \left| \omega \right|$, and hence $\chi_i^0 (\omega) \propto \omega^3$. 
The diverging behavior of the local density of states at $\omega=0$ in the noninteracting limit for model I is due to the fact that 
the numbers of sites (including the impurity site) on $A$ and $B$ sublattices, $N_A$ and $N_B$, respectively, 
are different for model I, but the same for models II and III, 
the similar discussion being given in the last part of Sec.~\ref{sec:model} for $\rho_0(\omega)$.

\begin{figure}[htbp]
\includegraphics[width=0.8\hsize]{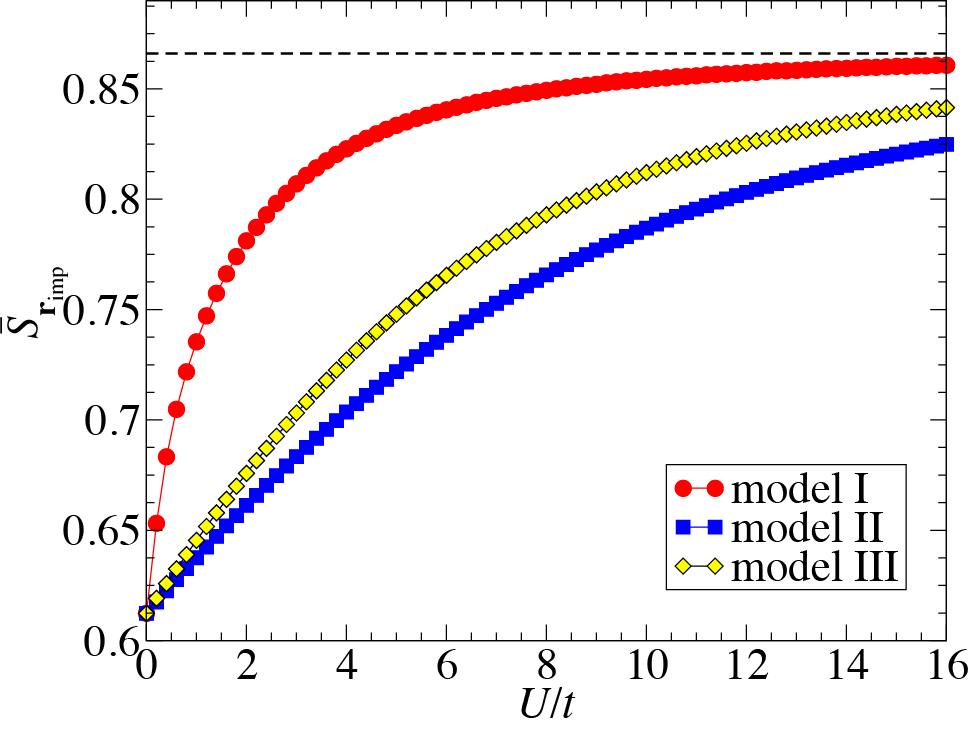}
\caption{(Color online) Local spin ${\bar S}_{{\bf r}_{\rm imp}}$ 
at the impurity site for $V=t$, $L=200$, and various values of $U$. For comparison, 
${\bar S}_{{\bf r}_{\rm imp}}$ in the strong coupling limit ($U\to\infty$) is indicated by dashed line. 
}
\label{fig:moment}
\end{figure}

The results of $\chi_i(\omega)$ calculated using the dynamical DMRG method for the three models are shown in 
Fig.~\ref{fig:msus}. 
First, it is noticed in Fig.~\ref{fig:msus} that the dynamical DMRG calculations well reproduce $\chi^0_i(\omega)$ obtained using Eq.~(\ref{eq:nonintchi}) 
with the same $L$ and $\eta$ for the noninteracting limit. 
In the case of finite interaction $U$, we find that $\chi_i (\omega)$ for model I diverges in the limit of $\omega \to 0$, 
which indicates the presence of free magnetic moment at the impurity site. 
Although the diverging behavior of $\chi_i(0)$ for finite $U$ seems similar to the one found in $\chi_i^0(0)$ for the noninteracting limit, 
we find in Fig.~\ref{fig:moment} that the local spin ${\bar S}_{{\bf r}_{\rm imp}}$ at the impurity site,
\begin{equation}
{\bar S}_{{\bf r}_{\rm imp}} = \sqrt{\langle \psi_0|{\bf S}_{{\bf r}_{\rm imp}}\cdot{\bf S}_{{\bf r}_{\rm imp}}|\psi_0\rangle},
\end{equation}
is sizably large for finite $U$ as compared to the one for the noninteracting limit. 
Here, the spin operator ${\bf S}_{\bf r}$ at site ${\bf r}$ is defined as
\begin{eqnarray}
\left({\bf S}_{\bf r}\right)_{\nu} = \frac{1}{2} \sum_{\sigma_1,\sigma_2} c_{{\bf r},\sigma_1}^{\dagger} 
\hat{ \sigma}_{\sigma_1,\sigma_2}^{\nu} c_{{\bf r},\sigma_2}  
\end{eqnarray}
and $\hat{\sigma}^\nu$ ($\nu= x,y,z$) is the $\nu$ component of Pauli matrices. 
In addition, as will be discussed later in Fig.~\ref{fig:specv100}, the local density of states at the impurity site is zero at the Fermi 
level for finite $U$, qualitatively different form the case for noninteracting limit. 
Therefore, we conclude that in the ground state of model I the local magnetic moment is not screened but rather isolated, 
and thus no Kondo screening occurs. 
This is in good accordance with the previous studies for the pseudogap Kondo 
problem~\cite{withoff,bulla2,gonzalez-baxton,ingersent,fritz2,vojta1,fritz1}.

On the other hand, as shown in Fig.~\ref{fig:msus}, $\chi_i (\omega)$ for models II and III monotonically decreases 
with decreasing $\omega$ for small $\omega$ and it becomes zero in the limit of $\omega \to 0$, 
which indicates the absence of free magnetic moment at the impurity site. 
Since $\lim_{\omega\to0}\chi_i^0 (\omega) \to 0$ already in the noninteracting limit for models II and III, 
the absence of free magnetic moment for a small $U$ region is related to the formation of bonding orbital composed of the 
impurity site and the surrounding conduction sites. 
However, as shown in Fig.~\ref{fig:moment}, the local spin ${\bar S}_{{\bf r}_{\rm imp}}$ at impurity site 
indeed increases with increasing $U$ smoothly to the strong coupling limit (i.e., $U\to\infty$), where a single electron is 
completely localized at the impurity site and only the spin degree of freedom is left. 
Therefore, these results imply that there is the crossover from a small $U$ region to a large $U$ region where the 
screening mechanisms are different: for a small $U$ region, the absence of free magnetic moment is due to 
the formation of bonding orbital, whereas for a large $U$ region the local magnetic moment is screened by the surrounding 
conduction electrons, i.e., the formation of a Kondo singlet state~\cite{krishna-murthy}.

Other noticeable effects of $U$ on $\chi_i(\omega)$ are summarized as follows. 
First, the line shape of $\chi_i (\omega)$ changes systematically with increasing $U$: 
the overall weight moves downward to a lower energy region with increasing $U$. 
This is associated 
with the decrease of the effective exchange interaction between the impurity site and the conduction site 
with increasing $U$ in the strong coupling limit. Second, the total spectral weight increases with $U$. 
Notice that the total spectral weight is related to the local spin ${\bar S}_{{\bf r}_{\rm imp}}$ at the impurity site, i.e., 
$ \int_0^\infty \chi_i (\omega){\rm d}\omega = {{\bar S}_{{\bf r}_{\rm imp}}}^2/3$. 
The larger $U$ increases the tendency of single occupancy at the impurity site with less charge fluctuations, 
which in turn increases the local magnetic moment, as seen in Fig.~\ref{fig:moment}.

\subsubsection{\label{sec:ldos}Local density of states at the impurity site}

The local density of states $\rho ({\bf r},\omega ) $ at site ${\bf r}$ is defined as 
\begin{eqnarray}
\rho ({\bf r},\omega ) = \left\{ 
\begin{array}{>{\displaystyle}c>{\displaystyle}c}
- \frac{1}{\pi} {\rm Im} G_{\rm e} ({\bf r},\omega+i\eta) & {\rm for}\ \omega > 0 \\
- \frac{1}{\pi} {\rm Im} G_{\rm h} ({\bf r},\omega+i\eta) & {\rm for}\ \omega <  0 
\end{array}
\right. 
\end{eqnarray}
where $G_{\rm e} ({\bf r},z)$ and $G_{\rm h} ({\bf r},z)$ are  
\begin{equation}
G_{\rm e} ({\bf r},z)= \left< \psi_0 \right| 
c_{{\bf r}, \sigma} (z - \mathcal{H}_\Gamma + E_0)^{-1} c_{{\bf r}, \sigma}^{\dagger} \left| \psi_0 \right>  
\end{equation}
and 
\begin{equation}
G_{\rm h} ({\bf r},z)= \left< \psi_0 \right| 
c_{{\bf r}, \sigma}^{\dagger} (z + \mathcal{H}_\Gamma - E_0)^{-1} c_{{\bf r}, \sigma} \left| \psi_0 \right>,  
\end{equation}
respectively. 
The local density of states $\rho_i (\omega )$ at the impurity site ${\bf r}_{\rm imp}$ is thus 
$\rho_i (\omega )= \rho ({\bf r}_{\rm imp},\omega)$.

\begin{figure}[htbp]
\begin{center}
\includegraphics[width=\hsize]{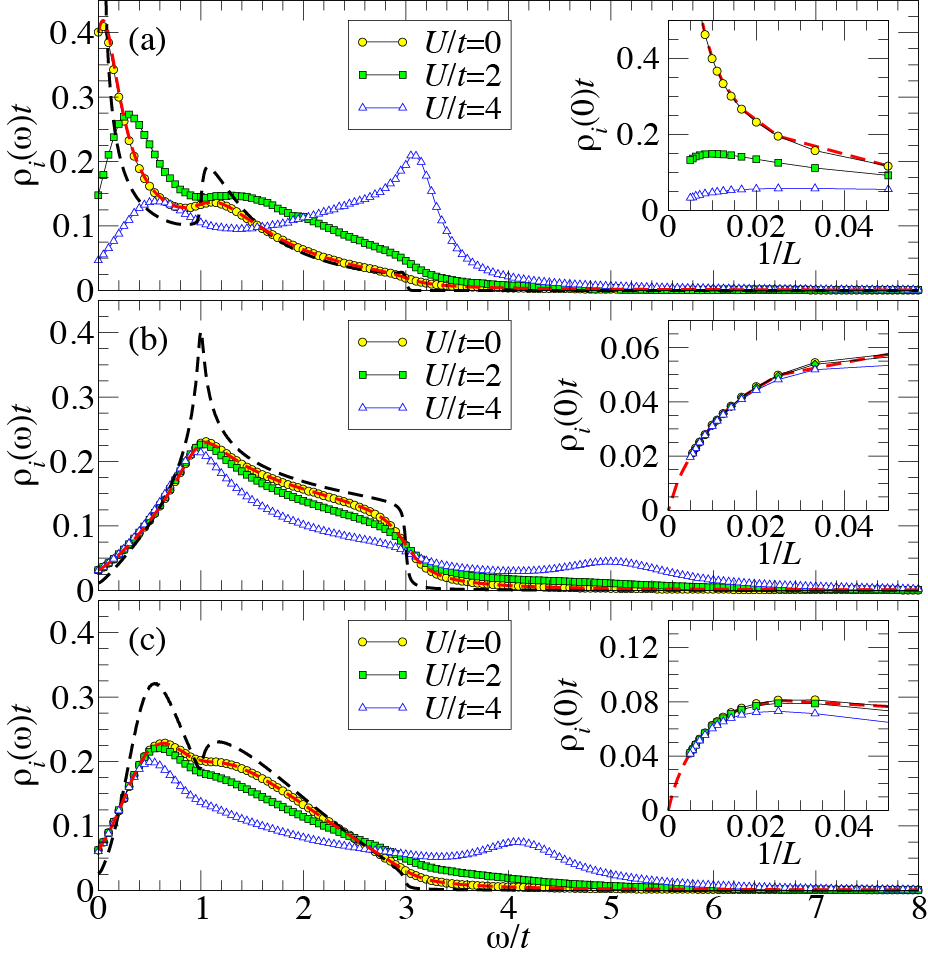}
\caption{(Color online) Local density of states $\rho_i (\omega)$ at the impurity site calculated using the dynamical 
DMRG method for models (a) I, (b) II, and (c) III. 
The parameters used are $L=100$, $V=t$, and $\eta = 20 t/L$ for $U/t=0$ (circles), $2$ (squares), and $4$ (triangles). 
For comparison, $\rho_i^0(\omega)$ for the noninteracting limit calculated using Eq.~(\ref{eq:nonintrho}) is also shown 
in red (black) dashed lines with $L=100$ (1000) and $\eta = 20 t/L$. 
Insets: $L$ dependence of $\rho_i (0)$ with keeping $\eta = 20 t/L$. For comparison, $\rho_i^0(\omega)$ calculated using 
Eq.~(\ref{eq:nonintrho}) is also plotted by red dashed lines. 
}
\label{fig:specv100}
\end{center}
\end{figure}

Figure~\ref{fig:specv100} shows the results of $\rho_i (\omega)$ for the three models calculated using the dynamical 
DMRG method. 
Since these models are particle-hole symmetric at half filling and the spectra are symmetric at $\omega = 0 $, we 
show $\rho_i (\omega)$ only for $\omega \ge 0$ in Fig.~\ref{fig:specv100}. 
For comparison, we also calculate the local density of states $\rho_i^0 (\omega )$ at the impurity site for the noninteracting 
limit by numerically diagonalizing the one-body part of the Hamiltonian ${\mathcal H}_\Gamma$ described 
by the Lanczos bases ${\bf a}_{\sigma}$ as in $\hat{H}_0^{\rm BL}$ in Eq.~(\ref{eq:modonebodyham}), i.e.,
\begin{eqnarray}
\rho_i^0 (\omega) = \frac{\eta}{\pi} \sum_{k \in (\varepsilon_{k} > \mu)} 
\frac{ \left| u_{{\bf r}_{\rm imp}}^{(k)} \right|^2 }{(\omega - \varepsilon_{k})^2 + \eta^2} 
\label{eq:nonintrho}
\end{eqnarray}
for $\omega \ge 0$. As shown in Fig.~\ref{fig:specv100}, the dynamical DMRG calculations well reproduce $\rho_i^0 (\omega )$ 
obtained using Eq.~(\ref{eq:nonintrho}) with the same $L$ and $\eta$.

Let us first focus on $\rho_i (\omega)$ for model I. As shown in Fig.~\ref{fig:specv100}(a), 
the spectral weight is redistributed drastically with increasing $U$.  
The diverging behavior of $\rho_i (\omega)$ in the limit of $\omega \to 0$ for $U=0$ is strongly suppressed 
and the low energy spectral weight is transferred to a higher energy region with increasing $U$. 
As shown in the inset of Fig.~\ref{fig:specv100}(a), we find that $\rho_i (0)$ for finite $U$ approaches to zero in the limit of  $L \to \infty$. 
This implies that for model I a small $U$ region is qualitatively different from the noninteracting limit but rather smoothly connected 
to the strong coupling limit where the charge fluctuations are completely suppressed and only the spin degree of freedom is left 
at the impurity site.

It is also observed in Fig.~\ref{fig:specv100}(a) that 
the lowest peak in $\rho_i (\omega)$ at $\omega / t \sim 0.3\, (0.5)$ for $U/t=2\, (4)$ 
becomes broader and the peak position shifts slightly to higher energy as $U$ increases. 
This is indeed consistent with the previous study of the same model using the QMC method~\cite{hu}. 
In addition, we find that with increasing $U$ the spectral weight in a much higher energy region is enhanced and 
gradually forms a peak structure, e.g, at $\omega / t \sim 3$ for $U/t=4$.

We shall next examine $\rho_i (\omega)$ for models II and III. 
As shown in Figs.~\ref{fig:specv100}(b) and \ref{fig:specv100}(c), we find that (i) 
$\rho_i (\omega)$ for the low energy region of $\omega/t\alt0.5$ is almost insensitive to the values of $U$ and (ii) 
$\rho_i (0)$ clearly becomes 0 in the limit of $L \to \infty$, thus exhibiting a pseudogap structure similar to the one for 
the noninteracting limit. 
The pseudogap structure in $\rho_i (\omega)$ is also found even for much larger $U$ (not shown). 
The fact that $\rho_i (\omega)$ for the low energy region is insensitive to $U$ is in good qualitative agreement with 
the previous study by the perturbation theory 
for the conventional Anderson impurity model, in which $\rho_i (\omega)$ at $\omega \sim 0$ for finite $U$ 
remains the same as the one for $U=0$~\cite{yamada,hewson}.

We also find in Figs.~\ref{fig:specv100}(b) and \ref{fig:specv100}(c) that the spectral weight in the high-energy region of 
$\omega/t>3$ increases 
with $U$, which is transferred from the low-energy region below $\sim3t$. This spectral weight redistribution with increasing $U$ is 
also very similar to the one in 
the conventional Anderson impurity model~\cite{yamada,hewson}, 
where the spectral weight in the low-energy region is suppressed and the high-energy peaks, corresponding to the lower and upper 
Hubbard peaks at $\omega\sim \pm U/2$, gradually emerge with increasing $U$, although the excitation energy of the high-energy peak 
found in Figs.~\ref{fig:specv100}(b) and \ref{fig:specv100}(c) is significantly different from $ U/2$. 
Therefore, $U$ dependence of the spectral weight for models II and III  can be qualitatively explained by the 
conventional Anderson impurity picture, except for the absence of Kondo resonance peak, which is simply due to the pseudogap 
structure in $\rho^0_i(\omega)$ at $\omega\sim 0$ for the noninteracting limit.

\subsubsection{\label{sec:sscf}Spin-spin correlation functions between the impurity site and the conduction sites}

Finally, we shall calculate the spin-spin correlation functions $S_{i} ({\bf r})$ between the magnetic impurity site at ${\bf r}_{\rm imp}$ 
and the conduction site at ${\bf r}$, 
\begin{eqnarray}
S_{i} ({\bf r}) = \left< \psi_0 \right| {\bf S}_{{\bf r}_{\rm imp}} \cdot {\bf S}_{{\bf r}} \left| \psi_0 \right>. 
\label{eq:sisc}
\end{eqnarray}
As described in Sec.~\ref{sec:orbs}, the ladder model is constructed separately for each conduction site 
${\bf r}$ to which the spin-spin correlation function $S_{i}({\bf r})$ is evaluated.

We should first note that one can easily construct the symmetric and antisymmetric BL bases for model II 
because the symmetry of the lattice structure remains the same as the one for the honeycomb lattice. 
However, it is not straightforward to construct the symmetric and antisymmetric BL bases for models I and III. 
In the case of model I, we can in principle perform the BL iterations taking as the initial BL bases two conduction sites, 
i.e., the conduction site of interest and the conduction site connected to the impurity site via $V$ 
[the conduction site denoted by cyan sphere in Fig.~\ref{fig:gmodel} (a)]. After the BL iterations are completed, 
the impurity site can be added to the resulting ladder model. 
With this slightly modified implementation, we can readily construct the symmetry adapted BL bases and the resulting ladder 
model is essentially decoupled for the symmetric and antisymmetric BL bases, as explained in Sec.~\ref{sec:sym}. 
However, this implementation naturally introduces an odd number of sites, which is problematic in the DMRG calculations. 
In the case of model III, it is generally difficult to construct the symmetry adapted BL bases. 
Therefore, we use the reduction scheme for spin degrees of freedom described in Sec.~\ref{sec:sym} 
(see also Fig.~\ref{fig:smap}) for models I and III.

\begin{figure}[htbp]
\begin{center}
\includegraphics[width=0.95\hsize]{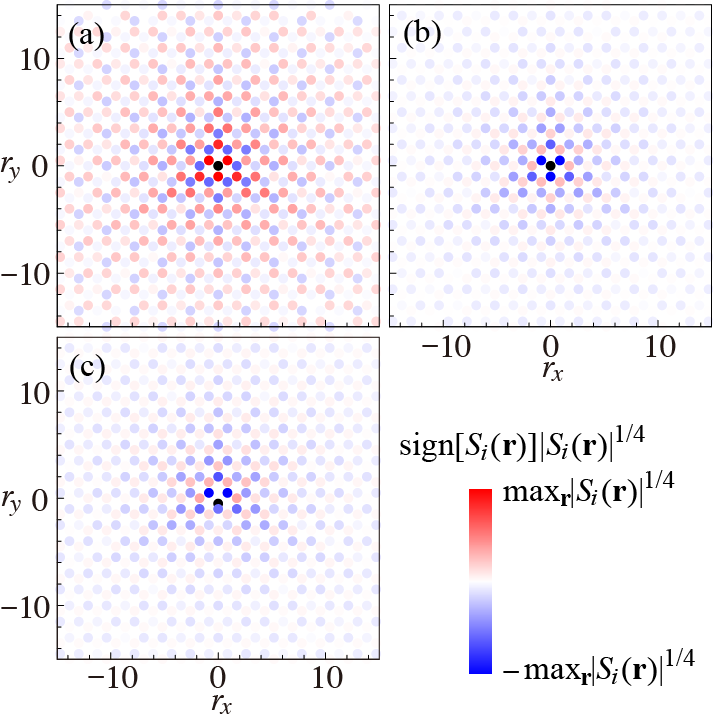}
\caption{(Color online) Intensity plot of the spin-spin correlation functions $S_{i}({\bf r})$ 
between the impurity site and the conduction sites for models (a) I, (b) II, and (c) III. 
The impurity site is located at ${\bf r}_{\rm imp} = (r_x,r_y)= (0,0)$ for (a) and (b), and 
${\bf r}_{\rm imp} = (0,-0.5)$ for (c), indicated by black circles. 
The parameters used are $U/t = 4$ and $V/t = 1$. 
The system size $L$ is chosen to satisfy $L=l_{\rm path}+100+{\rm mod}(l_{\rm path},2)$ 
where $l_{\rm path}$ is the minimum path length to reach the conduction site from the impurity site in the 
honeycomb lattice. 
Notice that the color intensity used here is for 
$\left| S_{i}({\bf r}) \right|^{1/4} {\rm sign}\left[S_{i}({\bf r})\right] $, instead of $S_{i}({\bf r})$ itself, for clarity. 
}
\label{fig:siscrs}
\end{center}
\end{figure}

Figure~\ref{fig:siscrs} shows the spatial distribution of the spin-spin correlation functions 
$S_{i} ({\bf r})$ for the three models. 
Because of the bipartite nature of the honeycomb lattice and the particle-hole symmetry at half filling, 
we can clearly see in Fig.~\ref{fig:siscrs} the alternating dependence of the sign of $S_{i}({\bf r})$ for all models: 
the spin-spin correlation functions $S_{i}({\bf r})$ at the conduction site belonging to 
the same (different) sublattice of the impurity site is positive (negative). 
We can also notice in Fig.~\ref{fig:siscrs} that model I exhibits relatively strong ferromagnetic correlations, while 
antiferromagnetic correlations are dominant for models II and III. 
The different behavior among the models is attributed to the fact that the ground state of model I is characterized 
with the appearance of unscreened local magnetic moment 
but the ground states of models II and III are instead both spin singlet, 
as discussed above in Sec~\ref{sec:dssf} and Sec.~\ref{sec:ldos}.

\begin{figure}[htbp]
\begin{center}
\includegraphics[width=0.75\hsize]{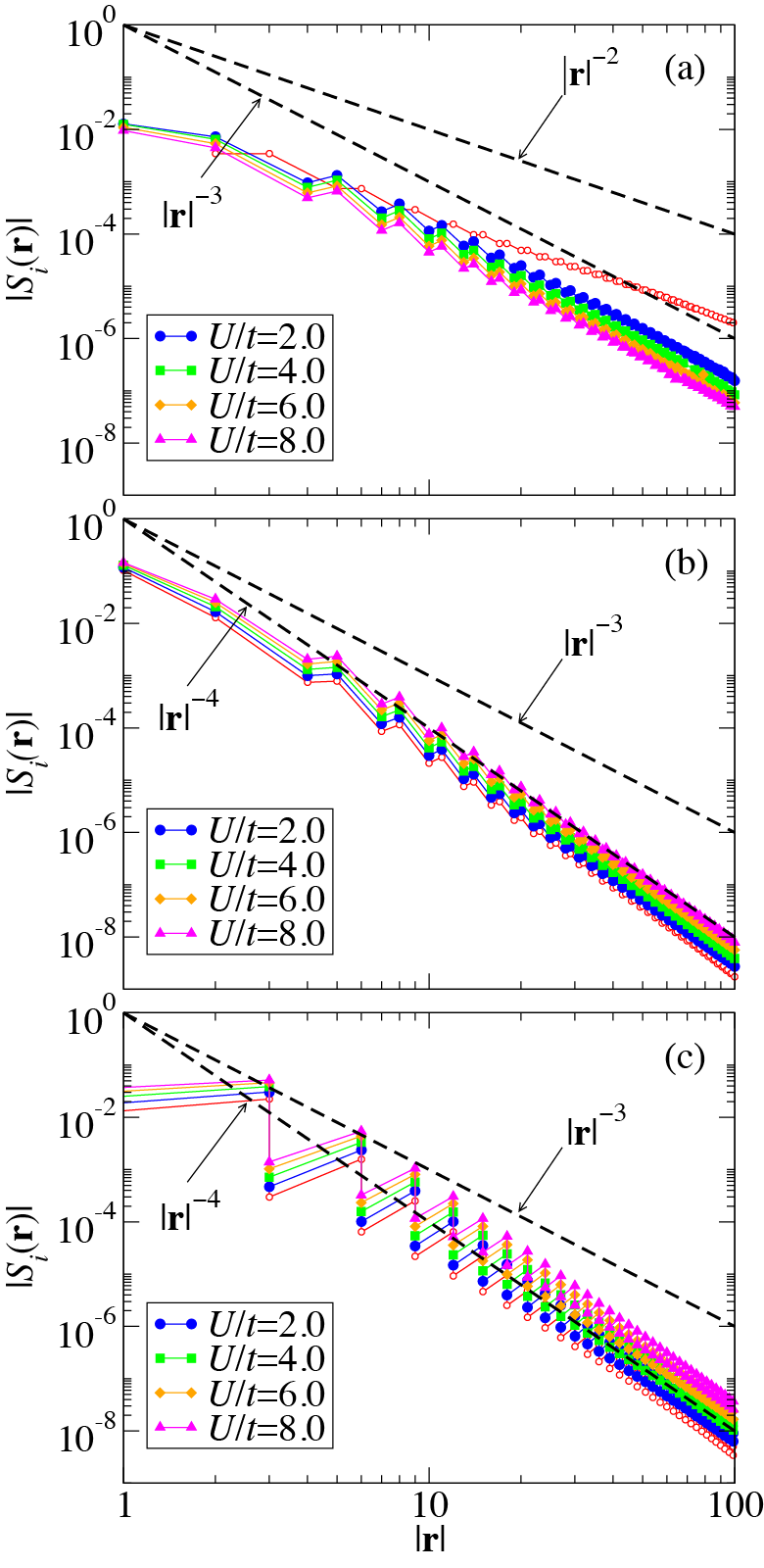}
\caption{(Color online) 
Log-log scale plots of the spin-spin correlation functions $S_{i}({\bf r})$ between the impurity site and the conduction sites 
for models (a) I, (b) II, and (c) III. 
The impurity site is located at ${\bf r}_{\rm imp} = (x,y)= (0,0)$ and the conduction sites ${\bf r}$ are chosen 
along $(0,1)$ direction (see Fig.~\ref{fig:siscrs}). 
The parameters used are $V/t = 1$ and different values of $U/t$ indicated in the figures. 
The system size $L$ is chosen to satisfy $L=l_{\rm path}+100+{\rm mod}(l_{\rm path},2)$ 
where $l_{\rm path}$ is the minimum path length to reach the conduction site from the impurity site in the honeycomb lattice. 
The spin-spin correlation functions $S_{i}^0({\bf r})$ for the noninteracting limit calculated using 
Eq.~(\ref{eq:nonintsisc}) are shown by red open circles. 
For comparison, $|{\bf r}|^{-\alpha}$ with different exponent $\alpha$ is also plotted by 
black dashed lines. 
}
\label{fig:siscdis}
\end{center}
\end{figure}

To discuss more details of the spin structures around the impurity site, the log-log scale plots of the spin-spin correlation 
functions $S_{i}({\bf r})$ for the three models are shown in Fig.~\ref{fig:siscdis}. We should notice first that the correlation 
functions can be very small for large $|{\bf r}|$, as small as $\sim 10^{-9}$--$10^{-8}$ at the maximum distance studied 
in Fig.~\ref{fig:siscdis}. 
However, we can still distinguish clearly the significant difference in the asymptotic behavior of $S_{i}({\bf r})$ for these models. 
We find in Fig.~\ref{fig:siscdis} that the spin-spin correlation functions $S_{i}({\bf r}) $ between the impurity site and 
the conduction sites decay as 
\begin{eqnarray}
  S_{i}({\bf r}) \propto  
  \left\{
    \begin{array}{cl}
      1/\left| {\bf r} \right|^{3}    &   \    {\rm for\,\, model\,\, I}  \\
      1/\left| {\bf r} \right|^{4}     &   \    {\rm for\,\, models\,\, II\,\, and\,\, III} 
    \end{array}
  \right.
\end{eqnarray}
in the asymptotic $|{\bf r}|$. 
These calculations thus demonstrate the capability of the BL-DMRG method to study spatially dependent quantities 
with extremely high accuracy.

To better understand these results, we also calculate the spin-spin correlation function $S_i^0({\bf r})$ for the noninteracting limit, 
which is given as 
\begin{eqnarray}
S_{i}^0({\bf r}) = \frac{3}{2} \sum_{k \in (\varepsilon_{k} < \mu)} \sum_{k^{\prime} \in (\varepsilon_{k^{\prime}} > \mu)} 
 ( u_{{\bf r}_{\rm imp}}^{(k)})^{\ast} u_{{\bf r}_{\rm imp}}^{(k^{\prime})} ( u_{\bf r}^{(k^{\prime})})^{\ast} u_{\bf r}^{(k)}. \nonumber \\
\label{eq:nonintsisc}
\end{eqnarray}
In the noninteracting limit, the spin-spin correlation functions between any two sites on the same sublattice 
are exactly zero whereas they are negative between any two sites on the different sublattices. 
As shown in Fig.~\ref{fig:siscdis}, we find that in the noninteracting limit the spin-spin correlation functions decay as 
\begin{eqnarray}
  S_{i}^0({\bf r}) \propto  
  \left\{
    \begin{array}{cl}
      1/\left| {\bf r} \right|^{2}    &   \    {\rm for\,\, model\,\, I}  \\
      1/\left| {\bf r} \right|^{4}     &   \    {\rm for\,\, models\,\, II\,\, and\,\, III} 
    \end{array}
  \right.
\end{eqnarray}
in the asymptotic $|{\bf r}|$. 
Therefore, the interaction $U$ drastically changes the exponent of the spin-spin correlation functions for model I. 
The asymptotic behavior of $S_{i}({\bf r})$ for model I with finite $U$ is rather the same as 
the one for Ruderman-Kittel-Kasuya-Yosida (RKKY) interaction~\cite{ruderman,kasuya,yosida} 
between two magnetic impurities coupled through the Dirac electrons on the honeycomb lattice at half filling, 
which has been indeed found to be as $ \propto \left| {\bf r} \right|^{-3}$~\cite{vozmediano,dugaev,brey,saremi,blackschaffer,sherafati,kogan}. 
In sharp contrast, the exponent remains the same for models II and III with and without $U$. 
The different effect of $U$ on the asymptotic behavior of $S_{i}({\bf r})$ for the three models is understood because 
the magnetic moment at the impurity site is not screened but rather isolated in the ground state of model I while 
the impurity moment is screened by the conduction electrons to form the spin singlet ground state for models II and III, 
as discussed in Sec.~\ref{sec:dssf} and Sec.~\ref{sec:ldos}.

It is also noticed in Fig.~\ref{fig:siscdis} that the absolute value of the spin-spin correlation functions are suppressed with increasing $U$ 
for model I, but they are enhanced for models II and III with positive (negative) values between the same (opposite) sublattices. 
These different behaviors are also understood by considering the different nature of the ground states of these models. 
The former results are due to the increase of unscreened local magnetic moment at the impurity site with increasing $U$ 
(see also Fig.~\ref{fig:moment}). 
The latter results are because the ground states for models II and III are both spin singlet, where the increased 
ferromagnetic correlations have to be compensated by enhancing the antiferromagnetic correlations.

\section{\label{sec:summary}Summary and Discussion}

We have introduced the BL-DMRG method for single- as well as multiple-impurity Anderson models in any spatial dimensions.  
The BL recursive technique is employed to map, without losing any geometrical information of the lattice, 
a general Anderson impurity model onto a Q1D model, to which the 
DMRG method can be applied with high accuracy. One of the key ideas in the BL-DMRG method is to include, 
as the initial BL bases, the Anderson impurity sites where the two-body interactions are finite. 
With this choice of the initial BL bases, the two-body interactions remain local in the resulting Q1D model. 
We have also introduced two reduction schemes to save the computational cost for the DMRG calculations. 
One is to construct the symmetry adapted BL bases when the Hamiltonian possesses a certain point group 
symmetry such as rotation and reflection. The other is to use spin degrees of freedom when the one-body part of the Hamiltonian is separated 
for up and down electrons. 
We have also discussed briefly the extension of the BL-DMRG method and the symmetry adapted BL bases for a multiorbital 
single-impurity Anderson model. 
Furthermore, we have demonstrated how the BL-DMRG method is applied to calculate spatially dependent quantities 
such as spin-spin correlation functions and local density of states at the conduction sites.

We should emphasize that the resulting Q1D model in the BL bases with $L$ sites along the leg direction 
represents the original Anderson impurity model in real space 
with approximately at least $\pi L^2$ and $4\pi L^3/3$ conduction sites in 
two and three spatial dimensions, respectively. Therefore, 
as long as the impurity properties are concerned, the BL-DMRG method can treat quite large systems for a wide class of 
Anderson impurity models, which are currently out of reach with the direct application of the QMC methods and the Lanczos exact diagonalization method. 
The spatially dependent quantities are rather difficult to calculated with the NRG method. Therefore, 
the BL-DMRG method has a great advantage on this aspect as well over the NRG method. 

As an application of the BL-DMRG method, we have studied the ground state properties 
of single-impurity Anderson models for graphene with an adatom and with a structural defect (vacancy). 
For this purpose, we have considered three different models: (i) a single impurity absorbed 
on the honeycomb lattice (model I), (ii) a substitutional impurity in the honeycomb lattice (model II), 
and (iii) an effective model for graphene with a single vacancy of carbon atom where the impurity site represents 
one of the $sp^2$ dangling orbitals at the carbon atoms surrounding the vacancy (model III). 
We have focused only on the particle-hole symmetric case at half filling and thus the electron density is always 
one, including at the impurity site. 
Our numerical results for the local magnetic susceptibility, the local spin, and the local density of states at the impurity site 
clearly show that the magnetic moment at the impurity site is not screened but rather isolated, 
and thus no Kondo screening occurs in the ground state of model I, 
while the impurity moment is screened by the conduction electrons to form the spin singlet ground state in models II and III. 

Moreover, we have applied the BL-DMRG method to calculate, with extremely high accuracy, 
the spin-spin correlation functions $S_i({\bf r})$ between the impurity site and the conduction sites for the three models. 
We have found the qualitative difference in the spatial distribution of the spin structures of the conduction electrons 
around the impurity site. The spin-spin correlation functions $S_i({\bf r})$ 
decay asymptotically as $\propto \left| {\bf r} \right|^{-3}$ for model I, the same asymptotic behavior 
as the one for the RKKY interaction between two magnetic impurities coupled to the Dirac conduction electrons, but qualitatively 
district from the one for the noninteracting limit ($\propto \left| {\bf r} \right|^{-2}$). 
On the other hand, the spin-spin correlation functions $S_i({\bf r})$ 
decay asymptotically as $\propto \left| {\bf r} \right|^{-4}$ for models II and III, which are exactly the same as the ones for the 
noninteracting limit. This difference can be understood because the magnetic moment in the ground state of model I is isolated 
but the spin singlet is formed in the ground state of models II and III. 

It is now interesting to discuss these results based on Lieb's theorem~\cite{lieb}. 
According to Lieb's theorem for bipartite lattice systems with no hopping between the same sublattices 
(except for the on-site potential), the total spin 
$S_{\rm tot}$ of the ground state at half filling is $S_{\rm tot}=\left| N_A - N_B \right|/2$, where $N_A$ ($N_B$) is the number 
of sites belonging to $A$ sublattice ($B$ sublattice)~\cite{lieb}. 
Regardless of the rigorous condition for Lieb's theorem~\cite{note1}, the theorem can be applied to the three models studied here 
because all models are bipartite and at half filling. Since model I has different number of sites (including the impurity site) 
on $A$ and $B$ sublattices, 
$\left| N_A - N_B \right| = 1$, the theorem predicts the total spin of the ground state is 1/2, which can be regarded as 
the isolated impurity spin. On the other hand, in models II and III, $N_A = N_B$ and thus the theorem predicts that the ground 
state of these models is spin singlet, which is also in accordance with our numerical results. 

We shall now discuss our results in comparison with the recent experiments on graphene. 
The experiments on graphene with hydrogen or fluorine adatoms as well as with structural defects (vacancies) have revealed that 
these systems carry magnetic moments with spin $1/2$ per adatom or vacancy and that these magnetic moments 
behave paramagnetically even at lowest temperatures~\cite{nair,maccreary}. 
Therefore, these experiments strongly indicate that no Kondo screening occurs. 
On the other hand, different experiments on graphene with vacancies have observed the Kondo-like signature in the temperature 
dependence of the resistivity~\cite{chen}. 
Although we have focused only on single impurity models with the particle-hole symmetry at half filling, our results 
should be relevant to these experiments as long as the number of adatoms or vacancies are dilute. 
We have found that the magnetic moment is unscreened but rather isolated in the ground state of model I, 
which therefore can explain, at least qualitatively, the spin $1/2$ free moment per adatom observed experimentally on graphene 
with hydrogen or fluorine adatoms~\cite{nair,maccreary}. 
On the other hand, we have found that the ground state of model III, a model for graphene with a single structural defect, 
is spin singlet and no free magnetic moment is found. 
Therefore, our results for model III are not in accordance with the experimental observation reported in Ref.~\cite{nair} but 
seem to be consistent qualitatively with experiments in Ref.~\cite{chen}. 

There are two comments regarding our results for model III and the experiments on graphene 
with vacancies reported in Ref.~\cite{chen}. 
First, it is reasonable that the impurity moment is screened to form the spin singlet ground state in model III. 
The reason is as follows. 
The number of electrons and the number of sites are both even in model III and therefore the ground state is closed shell in the 
noninteracting limit. Assuming the adiabatic evolution of the ground state with interaction $U$, the ground state must be total 
spin $S_{\rm tot}=0$ 
unless the correlation induces a level cross between the ground state and a low-lying excited state. 
In the experiments, the number of electrons removed by introducing vacancies must be even 
(i.e., six electrons removed per vacancy), and thus the system easily forms a close shell state with the total spin $S_{\rm tot}=0$ 
or possibly nonzero integer spin, but not with $S_{\rm tot}=1/2$. 
Second, although our results for model III seem to be consistent with the experimental observation in 
Ref~\cite{chen}, there is the following fundamental discrepancy. 
By controlling the number of electrons through gate voltage, it is found experimentally that the highest Kondo temperature appears 
away from half filling~\cite{chen}. 
This observation seems contradict to our calculations because the diverging hybridization function $\Delta(\omega)$ 
at $\omega = 0$ 
should induce the most tightly screened state and thus the highest Kondo temperature at half filling, but not away from half filling, 
for model III. 
The discrepancy between our results and the experiments as well as the disagreement between the two experiments suggest 
that the understanding of physics of graphene with vacancies and the corresponding 
magnetic properties would be beyond the simple model studied here and deserve further investigation both 
theoretically and experimentally.

Finally, we shall briefly comment on further possible extensions of the BL-DMRG method. 
The method is quite general and can be applied to general Anderson impurity models in any spatial dimensions. 
One major advantage of this method is its flexibility for the form of the conduction 
Hamiltonian. In this paper, we have studied Anderson impurity models in the real-space representation. 
However, the BL-DMRG method can be applied, without any difficulties, to Anderson impurity models in the energy-space 
representation (see Appendix \ref{app:hybfunc}).  
The BL-DMRG method in the energy-space representation allows us, in principle, to do the calculations in the thermodynamic limit 
once the hybridization function $\Delta(\omega)$ is evaluated accurately (see Appendix \ref{app:hybfunc}). 
The implementation of these extensions is straightforward and we believe that 
the BL-DMRG method in the energy-space representation should be valuable, e.g., 
for application as an impurity solver of DMFT for realistic electronic structure calculations of correlated materials~\cite{DFT+DMFT}. 
Research along this line is now in progress~\cite{shirakawa}. 

\section*{ACKNOWLEDGMENTS} 
The authors are grateful to H. Watanabe, E. Minamitani, and W. Ku for valuable discussion. 
The computation has been done using the RIKEN Cluster of Clusters (RICC). 
This work has been supported by Grant-in-Aid for Scientific Research from MEXT 
Japan under the Grant Nos. 24740269 and 26800171, and in part by RIKEN iTHES Project and Molecular Systems.

\appendix

\section{\label{app:hybfunc}The Hybridization Function of a General Anderson Impurity Model}

As mentioned in Sec.~\ref{sec:model}, the difference among different Anderson impurity models appears only through the hybridization 
function as long as the Anderson impurity terms are the same. Therefore, the hybridization function determines the physics of Anderson 
impurity models. In this Appendix, we shall derive the hybridization function for a 
general Anderson impurity model described by the Hamiltonian ${\mathcal H}_{\rm AIM}$ in Eq.~(\ref{eq:generalham}), 
and show that indeed the model difference appears through the hybridization function. 
The hybridization function is also required to apply 
the BL-DMRG method to Anderson impurity models in the energy-space representation.

To this end, we shall use the path integral formulation for a general Anderson impurity model ${\mathcal H}_{\rm AIM}$. 
The partition function $Z$ for ${\mathcal H}_{\rm AIM}$ is given as 
\begin{equation}
Z = \int \mathcal{D}d^{\ast} \mathcal{D}d \mathcal{D} c^{\ast} \mathcal{D} c 
\exp \left[ -S(d^{\ast},d,c^{\ast},c)\right], 
\label{eq:Z}
\end{equation}
where 
\begin{equation}
S(d^{\ast},d,c^{\ast},c) = S_0 (d^{\ast},d,c^{\ast},c)  + S_U (d^{\ast},d), 
\end{equation}
and 
\begin{eqnarray}
&{}& S_0(d^{\ast},d,c^{\ast},c) = - {\rm Tr}_{i\omega_n} \sum_{\sigma}
\left( {\bf d}_{\sigma}^{\dagger} (i\omega_n), {\bf c}_{\sigma}^{\dagger} (i\omega_n) \right) \nonumber \\
&{}& \quad \times 
\left(
\begin{array}{cc}
i\omega_n - \hat{H}_d & -\hat{V} \\
-\hat{V}^{\dagger} & i\omega_n - \hat{H}_c \\
\end{array}
\right) \left( 
\begin{array}{c}
{\bf d}_{\sigma}(i\omega_n) \\
{\bf c}_{\sigma}(i\omega_n) \\
\end{array}
\right). 
\end{eqnarray}
Here, $S_0 (d^{\ast},d,c^{\ast},c)$ [$S_U (d^{\ast},d)$] is the one-body part 
(the two-body part) of the total action $S(d^{\ast},d,c^{\ast},c)$, and 
\begin{equation}
{\bf d}^{\dagger}_{\sigma}(i\omega_n) = (d_{1,\sigma}^{\ast}(i\omega_n), d_{2,\sigma}^{\ast}(i\omega_n), \cdots, d_{M,\sigma}^{\ast}(i\omega_n) ) 
\end{equation}
and 
\begin{equation}
{\bf c}^{\dagger}_{\sigma}(i\omega_n) = (c_{1,\sigma}^{\ast}(i\omega_n), c_{2,\sigma}^{\ast}(i\omega_n), \cdots, c_{N,\sigma}^{\ast}(i\omega_n) ) 
\end{equation}
are Grassmann variables, corresponding to $d_{m,\sigma}^{\dagger}$ and $c_{n,\sigma}^{\dagger}$, respectively, at Matsubara 
frequency $i\omega_n$. 
$\mathrm{Tr}_{i\omega_n}$ indicates the sum over the Matsubara frequencies. The matrices ${\hat H}_d$, ${\hat H}_c$, and ${\hat V}$ are 
defined in Eq.~(\ref{eq:onebodyham}).

Carrying out the Gaussian integrals over variable $c^{\ast}$ and $c$ in Eq.~(\ref{eq:Z}), 
we obtain an effective action 
$S_{\rm eff}(d^{\ast},d)$ for variables $d^{\ast}$ and $d$, i.e.,  
\begin{equation}
S_{\rm eff}(d^{\ast},d) = S_0 (d^{\ast},d) + S_U (d^{\ast},d), 
\label{eq:Seff}
\end{equation}
where 
\begin{eqnarray}\label{eq:s_0}
S_0 (d^{\ast},d) = - {\rm Tr}_{i\omega_n} \sum_{\sigma} {\bf d}_{\sigma}^{\dagger}(i\omega_n) \nonumber \\
\times (i\omega_n - \hat{H}_d - \hat{\Gamma} (i\omega_n)) {\bf d}_{\sigma} (i\omega_n) 
\end{eqnarray}
and $\hat{\Gamma} (z)$ is the hybridization function for a complex frequency $z$ defined as 
\begin{eqnarray}
\hat{\Gamma} (z) = \hat{V} (z - \hat{H}_c )^{-1} \hat{V}^{\dagger}.
\label{eq:Delta}
\end{eqnarray}
To derive the above formula, 
we have used the following identity on Grassmann variables: 
\begin{eqnarray}
&{}& \int \prod_{i=1}^N {\rm d}x_i^{\ast} {\rm d}x_i 
\exp \left[ -{\bf x}^{\dagger} \hat{A} {\bf x} + {\bf x}^{\dagger} \hat{B}^{\dagger} {\bf y} + {\bf y}^{\dagger} \hat{B} {\bf x} \right] \nonumber \\
&=& {\rm det}(\hat{A}) \exp \left[ {\bf y}^{\dagger} \hat{B} \hat{A}^{-1} \hat{B}^{\dagger} {\bf y} \right],
\end{eqnarray}
where $\hat{A}$ is a regular $N \times N$ matrix, $\hat{B}$ is a $M \times N$ matrix, and 
\begin{eqnarray}
{\bf x}^{\dagger} = (x_1^{\ast}, x_2^{\ast}, \cdots, x_N^{\ast}), \\
{\bf y}^{\dagger} = (y_1^{\ast}, y_2^{\ast}, \cdots, y_M^{\ast}), 
\end{eqnarray}
are the vector representations for Grassmann variables $x_i^{\ast}$ and $y_i^{\ast}$, respectively.

It is now obvious from Eqs.~(\ref{eq:Seff}) and (\ref{eq:s_0}) that 
the effective action $S_{\rm eff}(d^{\ast},d)$ for the impurity sites depends on the conduction sites only 
through the hybridization function $\hat\Gamma(z)$. 
Therefore, all properties at the impurity sites are determined solely by ${\hat\Gamma}(z)$ 
when the Anderson impurity term ${\mathcal H}_d$  is the same. 
In other words, as long as the impurity 
properties are concerned, any models with the same ${\mathcal H}_d$ and ${\mathcal H}_U$ are equivalent if 
${\mathcal H}_c$ and ${\mathcal H}_V$ generates the same ${\hat\Gamma}(z)$. 
Therefore, we can even consider Eqs.~(\ref{eq:Seff}) and (\ref{eq:s_0}) as an effective Anderson impurity 
model in the complex-frequency representation 
which describes exactly the same physics of the original model ${\mathcal H}_{\rm AIM}$ in the real-space representation. 

Next, to derive the relation between the hybridization function $\hat \Gamma(z)$ for a complex frequency $z$ 
and the noninteracting Green's function, and also the recurrence relation for the noninteracting Green's function, 
we will use the following basic matrix algebra. 
Assuming that matrix $\hat X$ is a regular square matrix, 
\begin{eqnarray}
\hat X=\left(
\begin{array}{cc}
\hat{X}_{11} & \hat{X}_{12} \\
\hat{X}_{21} & \hat{X}_{22} \\
\end{array}
\right),
\end{eqnarray}
with $\hat X_{11}$ being a $r \times r$ matrix, the first $ r \times r$ elements $\hat Y_{11}$ of the inverse matrix of $\hat X$ is the inverse of 
the Schur complement of ${\hat X}_{22}$, i.e.,  
\begin{equation}
\hat{Y}_{11} = ( \hat{X}_{11} - \hat{X}_{12} \hat{X}_{22}^{-1} \hat{X}_{21} )^{-1},  
\label{eq:appidentity}
\end{equation}
where 
\begin{eqnarray}
\left(
\begin{array}{cc}
\hat{X}_{11} & \hat{X}_{12} \\
\hat{X}_{21} & \hat{X}_{22} \\
\end{array}
\right)
\left(
\begin{array}{cc}
\hat{Y}_{11} & \hat{Y}_{12} \\
\hat{Y}_{21} & \hat{Y}_{22} \\
\end{array}
\right) = \hat{1} 
\end{eqnarray}
and we assume that $\hat X_{22}$ is a regular matrix. 

We shall now derive the formula for the hybridization function $\hat{\Gamma}_{\rm BL}(z)$ 
in the BL bases ${\bf a}_\sigma^\dag$ and ${\bf a}_\sigma$ [Eq.~(\ref{a_op})], which block-tridiagonalize ${\hat H}_0$ 
in the form of $\hat{H}_0^{\rm BL}$, as shown in  Eq.~(\ref{eq:modonebodyham}). 
First, notice that since the noninteracting Green's function 
for a complex frequency $z$ is defined as 
\begin{eqnarray}
&{}& \hat{G} (z) =
\left(
\begin{array}{cc}
\hat{G}_{dd} (z) & \hat{G}_{dc} (z) \\
\hat{G}_{cd} (z) & \hat{G}_{cc} (z) \\
\end{array}
\right) \nonumber \\
&{}& \quad \quad \quad =
 \left(
\begin{array}{cc}
z - \hat{H}_d & -\hat{V} \\
-\hat{V}^{\dagger} & z - \hat{H}_c \\
\end{array}
\right)^{-1}
\end{eqnarray}
in the original conduction site bases ${\bf c}_\sigma^\dag$ and ${\bf c}_\sigma$ [Eq.~(\ref{eq:cop})], 
the impurity-site components of the Green's function, $\hat{G}_{dd} (z)$, is related to the hybridization function 
$\hat\Gamma(z)$ through   
\begin{equation}
\hat{\Gamma} (z) =  z - \hat{H}_d -\hat{G}_{dd}^{-1}(z).
\label{eq:d_g}
\end{equation} 

Next, we introduce the following matrix ${\hat G}^{(l)}_{\rm BL}(z)$ in the BL bases ${\bf a}_\sigma^\dag$ and ${\bf a}_\sigma$, 
defined as a part of the block matrices in $\hat{H}_0^{\rm BL}$: 
\begin{eqnarray}
&{}& \hat{G}^{(l)}_{\rm BL} (z) = \left( 
\begin{array}{ccc}
\hat{G}_{11}^{(l)}(z) & \hat{G}_{12}^{(l)} (z) & \cdots \\
\hat{G}_{21}^{(l)}(z) & \hat{G}_{22}^{(l)} (z) & \cdots \\
\vdots & \vdots & \ddots \\
\end{array}
\right) \nonumber \\
&=& \left( 
\begin{array}{cccc}
z - \hat{E}_{L-l} & -\hat{T}_{L-l} & \cdots & 0 \\
- \hat{T}_{L-l}^{\dagger} & z - \hat{E}_{L-l+1} & \cdots & 0 \\
\vdots & \vdots & \ddots & \vdots \\
0 & 0 & \cdots & z - \hat{E}_{L} \\
\end{array}
\right)^{-1}, 
\label{eq:gbl}
\end{eqnarray}
where $l=0,1,2,\cdots, L-1$ and ${\hat G}^{(0)}_{\rm BL} (z) =\hat{G}^{(0)}_{11}(z) = (z - \hat{E}_L)^{-1}$. 
The noninteracting Green's function is then expressed simply as $\hat{G}^{(L-1)}_{\rm BL} (z) $ 
in the BL bases obtained after the $L$-th BL iteration. 
Therefore, the hybridization function $\hat{\Gamma}_{\rm BL}(z)$ in the BL bases is given as 
\begin{eqnarray}
\hat{\Gamma}_{\rm BL}(z) =  z - \hat{E}_1 -\left [ \hat{G}_{11}^{(L-1)}(z) \right]^{-1}. \label{eq:appendix2}
\end{eqnarray} 

It is important to notice here that because of the block-tridiagonal form of the matrix $\hat{G}^{(l)}_{\rm BL} (z)$ in Eq.~(\ref{eq:gbl}), 
the following recurrence relation is satisfied: 
\begin{eqnarray}
\hat{G}_{11}^{(l)}(z) = \left( z - \hat{E}_{L-l} - \hat{T}_{L-l} \hat{G}_{11}^{(l-1)}(z) \hat{T}_{L-l}^{\dagger}
\right)^{-1}. \label{eq:appendix3}
\end{eqnarray}
This can be readily shown by using Eq.~(\ref{eq:appidentity}). 
Finally, using Eqs.~(\ref{eq:appendix2}) and (\ref{eq:appendix3}), 
we obtain the following form for the hybridization function $\hat{\Gamma}_{\rm BL} (z)$ for a complex frequency $z$ in the BL bases: 
\begin{eqnarray}
\hat{\Gamma}_{\rm BL} (z) &=& \hat{T}_1 \hat{G}_{11}^{(L-2)} (z) \nonumber \\
&=& \hat{T}_1 [z - \hat{E}_2 - \hat{T}_2 [ z - \hat{E}_3 - \cdots ]^{-1} \hat{T}_2^{\dagger} ]^{-1} \hat{T}_1^{\dagger}. 
\nonumber \\
\label{eq:drec}
\end{eqnarray}
Clearly, this is a matrix extension of the continued fraction formula~\cite{gagliano} 
and a similar formula has been used in the recursive Green's function technique~\cite{thouless,mackinnon,lewenkopf}. 
The recursive form for $\hat{\Gamma}_{\rm BL} (z)$ in Eq.~(\ref{eq:drec}) allows us to evaluate 
the hybridization function very accurately as compared to 
the simple full diagonalization method since the recursive method can treat much larger matrix sizes.

Now, recall that 
\begin{equation}
{\hat G}_{\rm BL}^{(L-1)}(z) = 
\arraycolsep5pt
\left(
\begin{array}{ccccc}
z-{\hat E}_1 & -{\hat T}_1 & 0 & \cdots &0\\
-{\hat T}_1^\dag &&&&\\
0&\multicolumn{4}{c}{\raisebox{-10pt}[0pt][0pt]{\large ${{\hat G}_{\rm BL}^{(L-2)}(z)}^{-1}$}}\\
\vdots&&&&\\
0&&&&\\
\end{array}
\right)^{-1}
\end{equation}
and therefore the matrix representation of the local density of states $\hat{\rho}_0 (\omega)$ for $\mathcal H_0$ at the second BL bases 
(i.e., at the sites next to the impurity sites in the Q1D model $\mathcal H_{\rm AIM}^{\rm Q1D}$) with $\hat{T}_1 = 0$ 
[see also Fig.~\ref{fig:1dmap}(b)] is 
\begin{eqnarray}
&{}& \hat{\rho}_0 (\omega) = 
- \frac{1}{\pi} \lim_{\delta \to 0^{+} } \left. \hat G_{22}^{(L-1)}(z) \right|_{z = \omega + i \delta,\ \hat T_1=0}\nonumber \\
&{}& = - \frac{1}{\pi} \lim_{\delta \to 0^{+} } \left. \hat G_{11}^{(L-2)}(z) \right|_{z = \omega + i \delta}\nonumber \\
&{}&= - \frac{1}{\pi} \lim_{\delta \to 0^{+} } \left. \left( 
z - \hat{E}_2 - \hat{T}_2 \left( z - \cdots \right)^{-1} \hat{T}_2^{\dagger}
\right)^{-1} \right|_{z = \omega + i \delta}. \nonumber \\
\end{eqnarray}
Here, $0^+$ is positive infinitesimal and we have used Eq.~(\ref{eq:appendix3}) in the third equality. 
Hence, we finally obtain the hybridization function $\hat{\Delta}(\omega)$ for a real frequency $\omega$ as
\begin{eqnarray}
\hat{\Delta}(\omega) &=& - {\rm Im} \hat{\Gamma}_{\rm BL}(\omega + i0^+) \nonumber \\
&=& \pi \hat{T}_1 \hat{\rho}_0 (\omega) \hat{T}_1^{\dagger}, 
\end{eqnarray}
where we have used Eq.~(\ref{eq:drec}). Since $\hat{\Gamma}(z)$ for a complex frequency $z$ is related to 
$\hat{\Delta}(\omega)$ for a real frequency $\omega$, 
\begin{eqnarray}
\hat{\Gamma}(z) = \frac{1}{\pi}\int_{-\infty}^{\infty} {\rm d} \omega \frac{1}{z - \omega} \hat{\Delta}(\omega),  
\end{eqnarray}
all properties at the impurity sites are determined by the hybridization function $\hat{\Delta}(\omega)$ for a real frequency $\omega$.

Now, consider the single-impurity Anderson models studied in Sec.~\ref{sec:model}. In this case, $M=1$ in $\mathcal H_{\rm AIM}$ 
and thus $\hat T_1$ and $\hat{\rho}_0 (\omega)$ are simply scalar. Therefore, 
we can readily find that $\hat{T}_1=V$ for model I, $\hat{T}_1=\sqrt{3}V$ for model II, and $\hat{T}_1=\sqrt{2}V$ for model III.


\begin{thebibliography}{99}
\bibitem{note5}
The first equality in Eq.~(\ref{eq:blbloc}) and Eq.~(\ref{eq:ut_c}) are valid only when ${\hat P}{\hat P}^\dag={\hat I}$. 
Therefore, when the BL iteration is terminated at the $L$th iteration ($L<N$), these equations are generally no longer true. However, 
even in this case, the second equality in Eq.~(\ref{eq:blbloc}) and Eq.~(\ref{eq:ut_a}) are satisfied.
\bibitem{note2}
When the number of the conduction sites (or orbitals) describe by $\mathcal{H}_c$ are infinite, the BL iteration should continue infinitely. 
However, in the practical calculations, 
$l$ is terminated at the size which can be treated by DMRG calculations, typically around few hundreds for the application 
in Sec.~\ref{sec:application}.
\bibitem{note3} 
Note that this reduction scheme can be applied similarly to a case where 
the one-body part of the Hamiltonian is separated for each component of 
Kramers doublets, e.g., when the relativistic spin-orbit coupling is finite.
\bibitem{note4}
It should be noted that the standard Lanczos recursive technique can be also applied to each irreducible representation separately  
to construct the Q1D model.
\bibitem{note1} The rigorous condition under which Lieb's theorem guarantees the nondegenerate and unique ground state, 
except for trivial degeneracy due to different values of total $S^z$, requires i) nonzero and positive on-site interactions at all 
sites and ii) even number of total sites. Since the three models studied here contains the on-site interaction only at the impurity site, 
Lieb's theorem does not guarantee the uniqueness of the ground state but simply predicts that one of the ground states, if degenerated, 
has the total spin $S=|N_A-N_B|/2$.
\end{thebibliography}
\end{document}